\documentclass{article}

\usepackage{arxiv}

\usepackage[utf8]{inputenc} 
\usepackage[T1]{fontenc}    
\usepackage{hyperref}       
\usepackage{url}            
\usepackage{booktabs}       
\usepackage{amsfonts}       
\usepackage{microtype}      
\usepackage{graphicx}
\usepackage{natbib}
\newcommand{\midtilde}{\raisebox{-0.20\baselineskip}{\textasciitilde}}
\title{Probabilistic Neural Network Tomography across Grane field (North Sea) from Surface Wave Dispersion Data}

\date{} 					

\author{
  Stephanie Earp \\
  Department of Geosciences\\
  University of Edinburgh\\
  \texttt{stephanie.earp@ed.ac.uk} \\
   \And
	Andrew Curtis\\
  Department of Geosciences\\
  University of Edinburgh\\
  and \\
    Institute of Geophysics\\
  ETH Zurich\\
    \texttt{andrew.curtis@ed.ac.uk} \\
   \And
	Xin Zhang\\
  Department of Geosciences\\
  University of Edinburgh\\
    \texttt{X.Zhang2@ed.ac.uk}  \\
   \And
	Fredrik Hansteen\\
  Equinor ASA \\
 Bergen, Norway\\
    \texttt{frhanst@equinor.com}      
}

\begin{document}
\maketitle

\begin{abstract}
Surface wave tomography uses measured dispersion properties of surface waves to infer the spatial distribution of subsurface properties such as shear-wave velocities. These properties can be estimated vertically below any geographical location at which surface wave dispersion data are available. As the inversion is significantly non-linear, Monte Carlo methods are often used to invert dispersion curves for shear-wave velocity profiles with depth to give a probabilistic solution. Such methods provide uncertainty information but are computationally expensive. Neural network based inversion provides a more efficient way to obtain probabilistic solutions when those solutions are required beneath many geographical locations. Unlike Monte Carlo methods, once a network has been trained it can be applied rapidly to perform any number of inversions. We train a class of neural networks called mixture density networks, to invert dispersion curves for shear-wave velocity models and their non-linearised uncertainty. Mixture density networks are able to produce fully probabilistic solutions in the form of weighted sums of multivariate analytic kernels such as  Gaussians, and we show that including data uncertainties in the mixture density network gives more reliable mean velocity estimates when data contains significant noise. The networks were applied to data from the Grane field in the Norwegian North sea to produce shear-wave velocity maps at several depth levels. Post-training we obtained probabilistic velocity profiles with depth beneath 26,772 locations to produce a 3D velocity model in 21 seconds on a standard desktop computer. This method is therefore ideally suited for rapid, repeated 3D subsurface imaging and monitoring.
\end{abstract}

\section{INTRODUCTION}
Seismic surface waves travel around the surface of the Earth but are sensitive to heterogeneity in elastic properties within the subsurface. Different frequencies of surface waves travel at different speeds since they depend mainly on the shear-wave velocity structure at different depths. Surface wave tomography uses this property (called dispersion) to infer the spatial distribution of subsurface shear velocities over global scales \citep{Woodhouse1984,Trampert1995,Shapiro2002,Zhou2006,Meier2007a,Meier2007}, regional scales \citep{Montagner1988,Curtis1997,Curtis1998,Ritzwoller1998,Devilee1999,Villasenor2001,Simons2002} and reservoir scales \citep{Bussat2011,deRidder2011,Mordret2014}.

Surface wave tomography is often performed using a 2-step inversion scheme \citep{Trampert1995,Ritzwoller2002}. In step 1, travel times of surface waves between pairs of known locations are measured at various fixed periods, then used to create geographical phase or group velocity maps at each period using 2D tomography. In step 2, the dispersion properties (speed of the waves at different periods -- often referred to as a dispersion curve) at each point on the 2D map are then inverted to estimate a 1D shear-wave velocity profile with depth below that point. The 1D velocity profiles beneath many geographical locations can then be placed side-by-side and interpolated to create a 3D model of the subsurface.

Both of the 2-step surface wave inverse problems are non-linear. They can be solved approximately by partially linearized \citep{Bodin2009}, or fully non-linear \citep{Rawlinson2014,Galetti2015,Galetti2016} Monte Carlo methods. These types of approaches provide relatively robust estimates of the range of possible shear wave velocity structures with depth that are consistent with the measured surface wave speeds (often referred to as the solution \textit{uncertainty}) by using the Markov chain Monte Carlo (McMC) algorithm to perform the inversions in a Bayesian framework. However, all existing sampling based methods, including the direct (1-step) 3D Monte Carlo tomography method of \cite{Zhang2018}, are extremely demanding computationally. If large data sets are to be inverted rapidly while maintaining our ability to assess post-inversion uncertainties without making linearizing approximations to the Physics, different methods are needed to speed up fully non-linear inversions.

We take an alternative approach and use neural networks to perform non-linear inversion of the phase velocities of Rayleigh-type Scholte surface waves (we refer to these simply as Rayleigh waves) for subsurface shear-wave velocity over length scales \midtilde 1-10km. Neural networks (NN's) approximate a non-linear mapping between two parameter spaces. The mapping is inferred from a set of examples of inputs and corresponding outputs of the real mapping (these examples are called \textit{training} data). Using certain types of NN-based methods, uncertainties in the mapping can be output by the network. NN's are therefore useful for problems where the forward mapping is well known or simple to calculate (in order to construct many training data) but the inverse mapping is complex or costly to determine directly. In such cases training data can be generated by applying the forward mapping to many sets of model parameter values, after which the NN can be trained to map in the inverse direction, taking the measurable data as input and outputting model parameter estimates.

Once trained, NN's can be applied to calculate the mapping for any input parameters extremely efficiently. For this reason neural networks have become increasingly popular for solving geophysical problems in recent years. Applications include well-log analysis \citep{Aristodemou2005,Maiti2007}, first arrival picking \citep{Murat1992,Mccormack1993}, fault detection \citep{Araya-Polo2017,Huang2017} and velocity analysis \citep{Roth1994,Calderon2000,Araya-Polo2018}. However all of these methods provide only deterministic estimates of the inverse problem solution (in most cases, the mean model estimate). Neural networks can also be used in a Bayesian sense to give fully probabilistic solutions. They were first used in Geophysics to estimate Bayesian uncertainties by \cite{Devilee1999} who inverted surface wave phase and group velocities for large-scale subsurface velocity structure and interface depths. They inverted regional dispersion curves for discretised probability distributions of crustal thickness across Eurasia using histogram and median networks, and analysed the trade-off between crustal thickness and velocity structure. \cite{Meier2007a,Meier2007} improved this method by using mixture density networks (MDN's) to give continuous probabilistic estimates of global crustal thickness and crustal velocity structure. A MDN is a type of network that maps an input vector to a probability density function (pdf) rather than to a single set of output values \citep{Bishop1995}. Since the work of \cite{Meier2007a,Meier2007}, mixture density networks have been used to perform petrophysical inversion of surface wave data for global water content and temperature in the mantle transition zone \citep{Meier2009}, inversion of industrial seismic data sets for subsurface porosity and clay content \citep{Shahraeeni2011,Shahraeeni2012}, inference of the Earth's 1D global average structure using body-wave travel-times \citep{DeWit2013} and for earthquake detection and source parameter estimation \citep{Kaufl2014,Kaufl2015}.

To produce a 3D shear-wave velocity versus depth model on any scale using the 2-step method, the inverse problem for structures with depth must be solved at many geographical locations (usually many thousands) over the area of interest. McMC inversion methods are computationally expensive and it is generally impractical to apply them in cases where parameters or data sets are large, where computational efficiency and processing time are usually limiting constraints due to the need to forward model many samples (of the order of thousands or millions) at each location. On the other hand, once trained, NNs and MDNs can often solve such inverse problems in seconds with no additional sampling. In addition, in cases where we wish to monitor changes in the subsurface, the same network can be applied rapidly to repeated data measurements, enabling the possibility of near-real time monitoring provided that the inputs to the networks can be produced rapidly from the raw measured data. Our aim herein is to investigate whether this is possible in practice.

In what follows we first introduce neural networks and mixture density networks and how they can be used to invert Rayleigh wave phase velocities for models of 1D shear-wave velocity with depth. We discuss the effect of data uncertainty and how to incorporate this within a neural network, then apply trained networks to field data from the Grane field in the Norwegian North Sea to create 2D shear-wave velocity maps of specific depth intervals. We compare the results from the MDN to non-linearized McMC methods, and thus prove that MDN surface wave inversion methods are both efficient and robust at the scale of reservoirs.
\section{METHOD}\label{method}
\subsection{Grane data}
\begin{figure}
	\begin{center}
	\includegraphics[width=\textwidth]{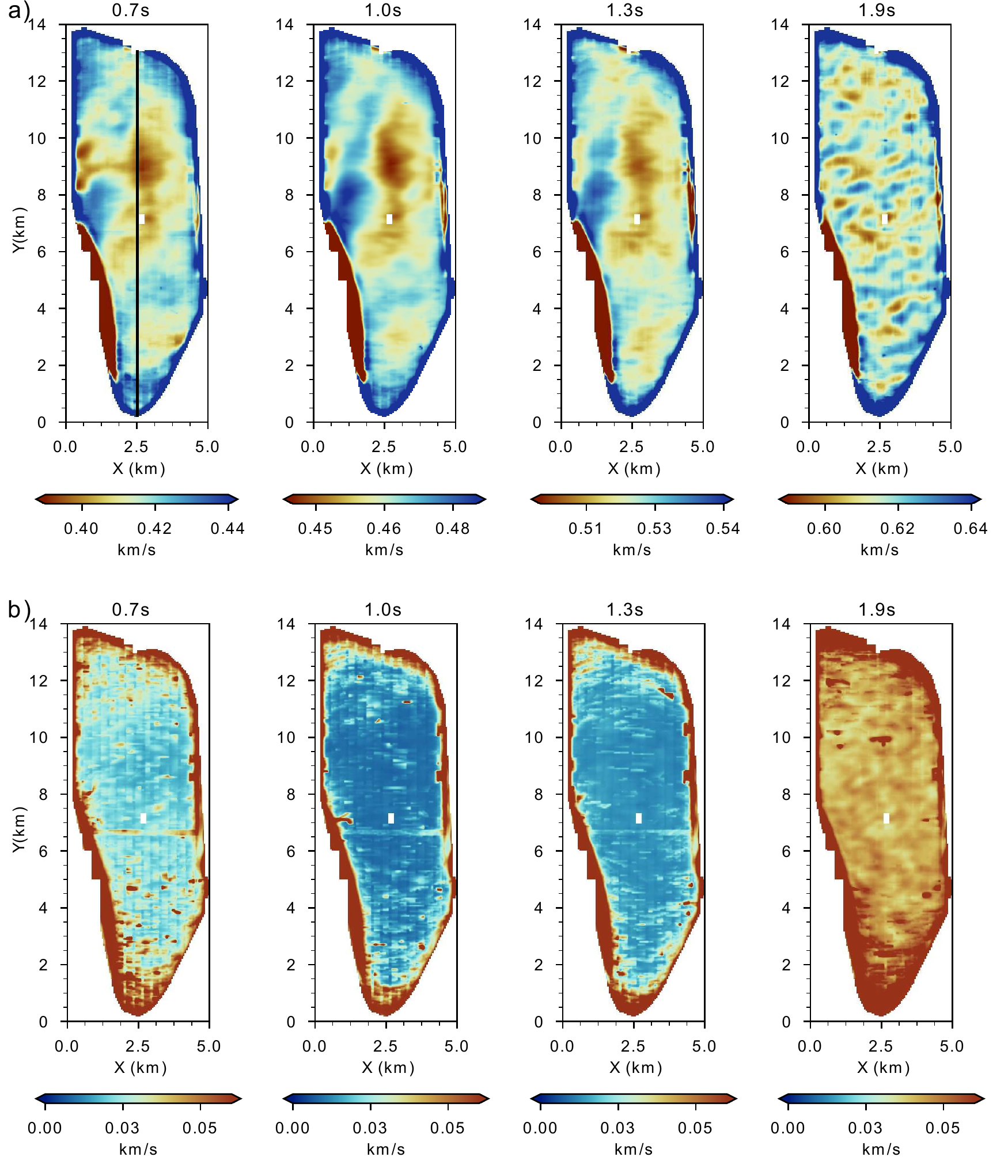}
	\end{center}
	\caption{(a) A selection of four phase velocity maps used to compute discretised dispersion curves. Periods shown are 0.7s, 1.0s, 1.3s and 1.9s.  (b) Maps of estimated standard deviation of uncertainties in the phase velocity at each location. Velocities and uncertainty colour scales are saturated at either end to prevent domination of outliers, and to highlight structure across the field. The vertical black line in the top left plot shows the location of a cross section shown in other Figures.}
	\label{fig:Phase_vel,Phase_uncer}
\end{figure}

Grane is an offshore oil field in the Norwegian North Sea. A permanent reservoir monitoring (PRM) system was installed in 2014 over approximately 50 km\textsuperscript{2} of the Grane seabed \citep{Thompson2015}. Ambient seismic noise is recorded continuously at the field using four-component sensors -- 3-component geophones (Vertical, North and East) and a hydrophone. The data used in this study was preprocessed according to the protocol of \cite{Zhang2019}, summarised as follows. Data from the vertical and hydrophone components were selected over a 6.5 hour interval. The data were bandpass-filtered between 0.35-1.5Hz and data from every pair of stations are cross-correlated using overlapping half-hour recording sections, then correlations are stacked over the full 6.5 hour interval. Cross-correlations of hydrophone and vertical component noise mainly contain information about Rayleigh-type waves. Phase velocities were automatically picked for the cross-correlation of each station-pair. Seventeen phase velocity maps and their corresponding standard deviation (uncertainty) maps were produced using eikonal tomography for periods between 0.6 to 2.2 seconds at 0.1 second intervals over a 50m x 50m grid. Figure \ref{fig:Phase_vel,Phase_uncer}a shows 4 examples of the phase velocity maps at periods 0.7s, 1.0s, 1.3s and 1.9s and their corresponding uncertainties.

\cite{Zhang2019} perform 1D, 2D and 3D Markov Chain Monte Carlo (McMC) Tomography over the Grane field to produce maps of the shear velocity structure with depth. However, McMC solutions are relatively slow to compute as they require \midtilde$10^6$ 3D or \midtilde$10^9$ 1D forward modelling simulations to obtain robust results and this needs to be repeated for each new data set for 4D surveys. Given that a PRM system exists, and that large amounts of data can be recorded in this area at many different points in time, a faster method is desired for monitoring applications.

\subsection{Bayesian Inference}
We wish to solve the surface wave inversion problem in a probabilistic framework to find the Bayesian posterior distribution of subsurface velocity structure parameters $\textbf{m}$ that fit the given data $\mathbf{d}$, written as $p(\textbf{m} \mid \textbf{d})$. This is defined as \citep[e.g.][]{Tarantola2005}:
\begin{equation}\label{Bayes}
p( \textbf{m}\mid \textbf{d}) = k\, p(\textbf{d} \mid \textbf{m}) \, p(\textbf{m})
\end{equation}
where $p(\textbf{m})$ represents the prior probability density on the velocity parameter space which describes information about $\textbf{m}$ known prior to using data $\textbf{d}$, $p(\textbf{d} \mid \textbf{m})$ is known as the likelihood and represents the conditional probability of measuring data $\textbf{d}$ given the velocity parameters $\textbf{m}$, and $k$ is a normalisation constant. In multidimensional problems where the dimensionality of $\textbf{m}$ is greater than 1, we often wish to infer the posterior inversion information about a single parameter with index $i$ and hence must calculate the marginal posterior distribution $p( m^i\mid \textbf{d})$. This is obtained by integrating over all parameters $m^j$ that are not of interest:
\begin{equation}\label{marg_post}
p( m^i\mid \textbf{d}) = \int_{\forall m^j\neq m^i} p(\textbf{m} \mid \textbf{d})\, dm^j
\end{equation}
In this study we focus on estimating marginal distributions $p( m^i\mid \textbf{d})$.
\subsection{Mixture Density Networks}
A neural network can be \textit{trained} to represent the arbitrary non-linear mapping between the spaces of input data $\mathbf{d}$ and output parameters $\mathbf{m}$ by presenting the network with a set of $N$ training pairs $T=\{(\mathbf{d}_i,\mathbf{m}_i) : i = 1,...,N\}$ and minimizing a cost function that measures the difference between the NN output and the defined output, often called the `error'. For example if the set of training velocity structures $\textbf{m}_i$ are distributed according to the prior posterior density function (pdf), then a network trained to output $\textbf{m}_i$ given input $\textbf{d}_i$ by minimizing the sum-of-squared errors across set $T$ will output an approximation to the mean of the Bayesian posterior distribution $p(\mathbf{m} \mid \mathbf{d)}$ when presented with data $\textbf{d}$ \citep{Bishop1995}. By contrast, in this paper we use a class of neural networks called mixture density networks. These provide a framework for modelling complete probability distributions. They are trained on the same set $T$ of data-velocity structure pairs, but instead of providing the mean estimate of the velocity structure, they provide an estimate of the Bayesian posterior probability distribution $p(\mathbf{m} \mid \mathbf{d)}$. The estimate is parametrised by a mixture (sum) of Gaussian kernels
\begin{equation}\label{MDN_mix}
p(\mathbf{m} \mid \mathbf{d)}=\sum_{k=1}^M\alpha_k\Theta_k(\mathbf{m} \mid \mathbf{d})
\end{equation}
where $\alpha_k$ are amplitude parameters that attach relative importance to each Gaussian kernel, $M$ is the number of Gaussians in the mixture, and $\Theta_k$ are Gaussian density functions given by
\begin{equation}\label{Gauss_mix}
\Theta_k(\textbf{m} \mid \textbf{d}) = \frac{1}{(2\pi)^{c/2}\sigma^c_k(\textbf{d})}exp\left\{-\frac{(m_k-\mu_k(\textbf{d}))^2}{2\sigma^2_k(\textbf{d})}\right\}
\end{equation}
where $c$ is the dimensionality of $\textbf{m}$.

The set of mixture parameters $\alpha_k$, means $\mu_k$ and standard deviations $\sigma_k$ fully define the set of Gaussian kernels and hence the output of the MDN. Training an MDN thus requires that we create a way to predict appropriate values for these parameters given any input data. For this task we use a standard feed-forward neural network which contains a set number of layers and nodes. At each layer the inputs of each node are weighted and summed before being passed through a function that induces non-linearity in the mapping. This provides an output value that can become the input for all units in the following layer. These weights are adjusted during training to provide the optimum mapping. The number of mixtures $M$ dictates the complexity of the final probability distribution, and the number of network outputs is given by $(c+2)\times M$ compared with the output of a standard NN that has $c$ outputs. The number of kernels that should be used depends on the complexity of the problem; however, as long as more kernels than necessary are used, an accurate number is not required as the network can reduce the amplitude of any mixture parameter to near zero for redundant kernels \citep{Bishop1995}. For a full description of MDN and neural network structures see \cite{Bishop1995}, or in a geophysical context see \cite{Meier2007} or \cite{Shahraeeni2011}.

During training the internal weights of the network are adjusted to maximize the likelihood of the desired probability density function given the training data. The cost function minimized is the negative log likelihood function \citep{Bishop1995}
\begin{equation}\label{MDN_cost}
E_{MDN}=-\sum_{n=1}^N ln\left\{\sum_{k=1}^M\alpha_k(d_n)\Theta_k(m_n\mid d_n)\right\}
\end{equation}
We train multiple networks with different network structures, and each network's parameters are randomly initialised before training begins. The networks are then combined using a weighted average of network outputs in order to improve generalisation and prediction accuracy \citep{Dietterich2000}. The weights for each network are determined by the cost function $E_{MDN}$ evaluated over the so-called test set -- a portion of the data set removed before training and used to test the network once training has completed. An approximation to the posterior probability distribution of a set of velocity parameters $\textbf{m}$ given data $\textbf{d}$ is thus given by
\begin{equation}\label{ensemble}
p(\textbf{m} \mid \textbf{d}) \simeq \sum_{k=1}^{L} \sum_{j=1}^{c} \frac{E_{k}\alpha_{kj}}{\sum^{L}_{l=1} E_l}\Theta_{kj}(\textbf{m}\mid\textbf{d})
\end{equation}
where 
\begin{equation}
E_k=-\exp(E_{MDN,k})
\end{equation}
and where $L$ is the number of networks in the ensemble, $E_{MDN,k}$ is the cost function value of the $k$th network, $\alpha_{kj}$ is the $j$th weighting parameter of the $k$th network, and $\Theta_{kj}$ is the $j$th Gaussian kernel of the $k$th network. Once the networks have been trained, the outputs can be used to estimate the posterior probability distributions using Equations \ref{Gauss_mix} and \ref{ensemble}. This gives a more complete description of the family of velocity structures that are consistent with the data than does the output of a standard NN.

\subsection{Creating a Training set}
\begin{figure}
	\begin{center}
		\includegraphics[width=0.6\textwidth]{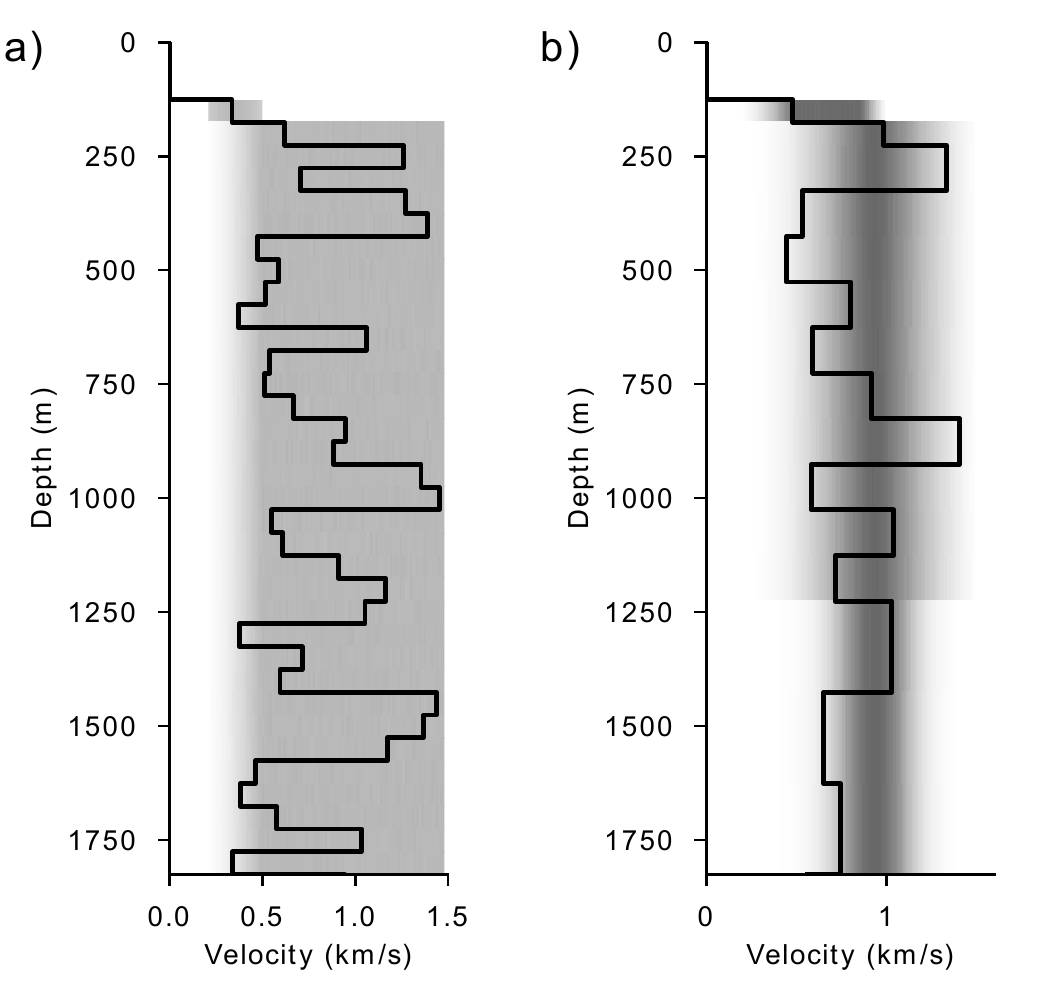}
	\end{center}
	\caption{(a) Initial distribution of velocity structures created with a piecewise-constant discretisation over depth. (b) Distribution of velocity structures created after averaging structures in (a) over larger depth intervals. Grey-scale shows the probability density distribution, darker colours represent higher density of velocity structures, and the black line is an example of a randomly selected velocity structure in each panel which also illustrates the depth intervals used in cases (a) and (b).}
	\label{fig:models_raw,models_sm}
\end{figure}

The velocity structures $\textbf{m}$ are parametrised as follows: each 1D structure has a water layer of 126m, followed by constant velocity layers every 25m to a depth of 100m below the water layer, then 50m thick layers down to 2000m below the water layer, beneath which there is a homogeneous half-space. For each velocity structure the S-wave velocity of the top solid layer was selected randomly from the Uniform probability distribution $v_{top} \sim U(0.2km/s,0.5km/s)$ to represent unconsolidated near-surface sediments. For a fundamental mode surface wave to be observed, the top solid layer must have the lowest velocity \citep{Galetti2016}, therefore the following layers were randomly selected from distribution $U(v_{top},1.5km/s)$. We generated 1,000,000 velocity structures and Figure \ref{fig:models_raw,models_sm}a shows the resulting distribution of velocities along with an example velocity structure.

The forward problem is solved for each of the generated velocity structures using the DISPER80 subroutines by \cite{Saito1988} to obtain corresponding fundamental mode Rayleigh wave dispersion curves. The phase velocities were calculated for periods 0.6-2.2s at 0.1s intervals in order to match the range available from ambient noise recorded at Grane. The DISPER80 forward modeller needs P and S-wave velocity and density for each layer in depth in order to calculate the phase velocities at each of any set of discrete periods. From our S-wave velocity structures calculated previously we computed a corresponding P-wave velocity $v_p$ and the density $\rho$ for each velocity layer based on typical values for sedimentary rocks using \citep{Castagna1985,Brocher2005}
\begin{eqnarray}
v_p = 1.16v_s+1.36 \\
\rho = 1.74v_p^{0.25}
\end{eqnarray}
Rather than attempt to invert surface wave speeds at 17 periods for shear velocities in 40 depth layers, before training the velocity model is averaged over seventeen larger fixed-depth intervals (Figure  \ref{fig:models_raw,models_sm}b). We then train networks to invert for the velocity in each of these larger depth intervals.

\subsection{Uncertainties}
In past work, uncertainty information about the data is only included by adding random Gaussian noise to the training data set \citep{Devilee1999,Meier2007a,Shahraeeni2011,DeWit2013}. Adding noise acts to regularise the network, helps to generalise when the network is inverting new data, and accounts for the data uncertainty in the Bayesian solution. However the disadvantage of such an approach is that when presenting the network with new data, updated uncertainty information for those particular data is not included in the inversion; indeed, that network would invert the data assuming that the incorrect (old) data uncertainties still pertain.

\begin{figure}
	\begin{center}
	\includegraphics[width=0.75\textwidth]{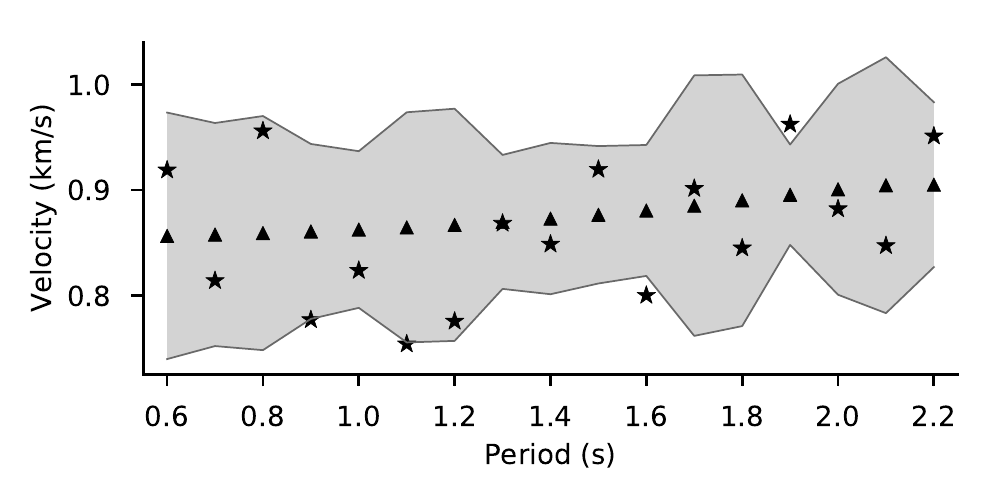}
	\end{center}
	\caption[Graph showing a synthetic dispersion curve $\mathbf{d}$ compared to a dispersion curve with added noise $ \tilde{\mathbf{d}}$]{Graph showing a synthetic dispersion curve $\mathbf{d}$ (triangles) compared to a dispersion curve with added noise $ \tilde{\mathbf{d}}$ (stars). The grey shaded area is the uncertainty $\mathbf{u}$ from Equation \ref{Uncer1}.}
	\label{fig:data_uncer}
\end{figure}

By contrast, here the data uncertainty is included as an additional set of inputs to the network. This makes sense because uncertainty is in fact additional pertinent information for each inversion. To train the MDN the clean synthetically-modelled data set is augmented with varying levels of noisy data. For each data point in the original synthetic data set a random percentage of noise $\epsilon$ is selected between the bounds outlined in Table \ref{Table_noise} for six different Uniform distributions of $\epsilon$. The noise is then added to the data according to
\begin{eqnarray}
u_j= \epsilon\times d_j \label{Uncer1}\\
\tilde{d}_j = \mathcal{N}(0,1) u_j + d_j \label{Uncer2}
\end{eqnarray}
where $u_j$ is the uncertainty value of the noisy data $\tilde{d}_j$ and $\mathcal{N}(0,1)$ is a Standard Normal distribution with mean 0 and standard deviation 1. An example of noisy data and the randomly chosen noise level is shown in Figure \ref{fig:data_uncer}. We thus generate an augmented training set of data-velocity structure pairs $T_{uncer} = \{([\tilde{\mathbf{d}}_j,\mathbf{u}_j],\mathbf{m}_j) : j = 1,...,N\}$, where our data consists of the noisy dispersion curves $\tilde{\mathbf{d}_j}$ and their associated uncertainties $\mathbf{u}_j$. The final data sets $T$ and $T_{uncer}$ are then shuffled and split into a training set (90\% of training pairs) that is used to train the network for the optimum mapping, a validation set (5\%) used during training to check the network is not over-fitting the training examples (see below), and a test set (5\%) which is used post training to assess the network performance on previously unseen data. This final assessment provides weights $E_i$ for the network ensemble in Equation \ref{ensemble}. Early stopping is employed to prevent over-fitting of the network to the data: this is where the cost function is periodically checked on the validation set during training. When the cost function stops decreasing it is assumed that the network is already fit to the training data but is no longer improving its generalisation to new data. Training is then stopped.

\begin{table}
\caption{Table of percentage range of noise added to data set in each of six different noise scenarios. Uncertainty is added to each datum according to Equations \ref{Uncer1} and \ref{Uncer2}. In the first scenario uncertainties are zero.}
\centering
\begin{tabular}{ccc} \toprule
 & \multicolumn{2}{c}{Percentage Noise \%}\\
Noise Scenario & Min & Max \\ \midrule
1 & 0 & 0 \\ \addlinespace[-0.5ex]
2 & 0 & 5 \\ \addlinespace[-0.5ex]
3 & 4 & 14 \\ \addlinespace[-0.5ex]
4 & 10 & 15 \\ \addlinespace[-0.5ex]
5 & 3 & 10 \\ \addlinespace[-0.5ex]
6 & 0 & 15 \\  \addlinespace[-0.5ex] \bottomrule
\end{tabular}
\label{Table_noise}
\end{table}

\section{RESULTS}\label{results}
\subsection{Network Design}
\begin{figure}
	\begin{center}
	\includegraphics[width=0.5\textwidth]{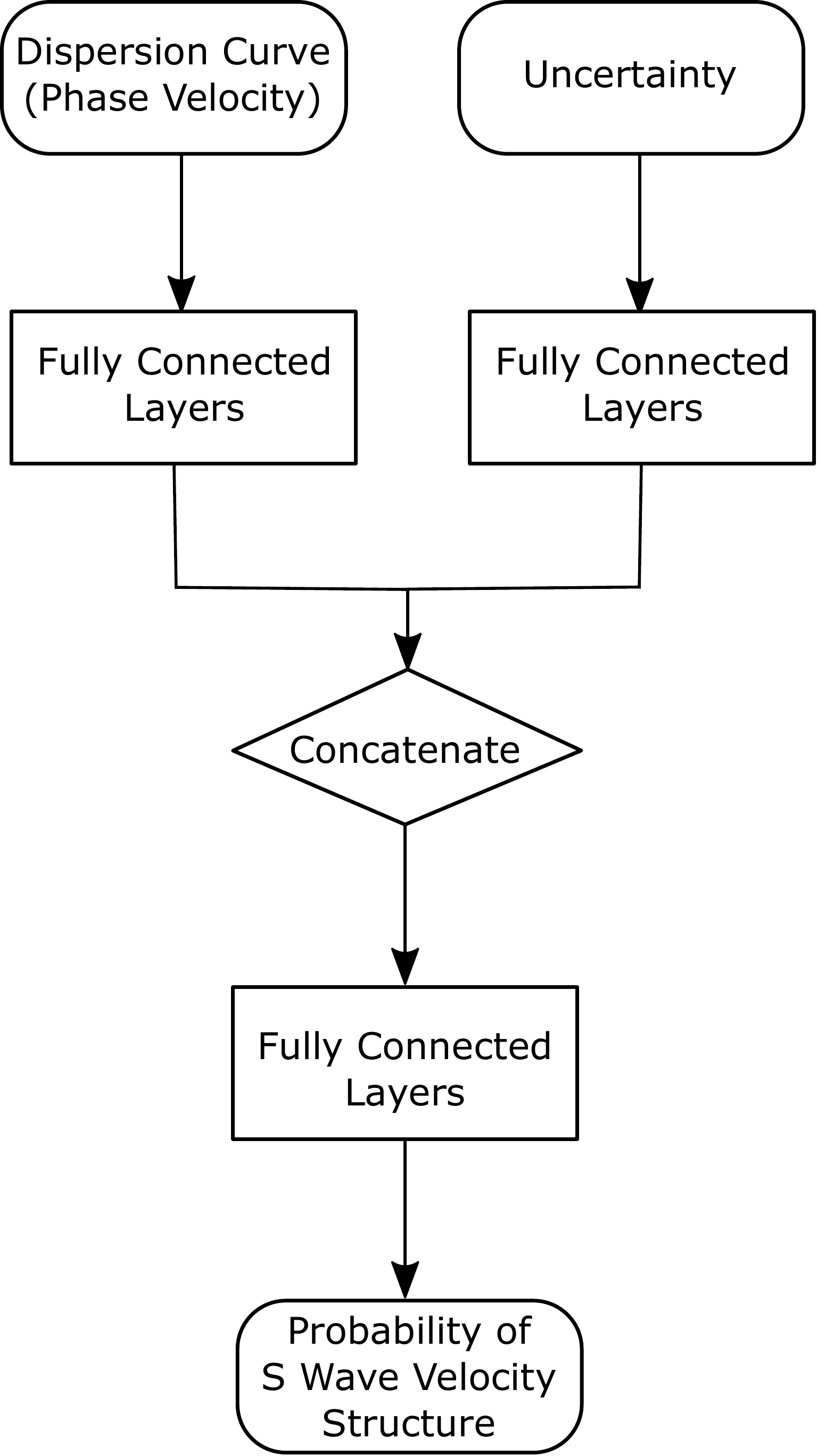}
	\end{center}
	\caption[Diagram of network used to include uncertainty estimates in the input vector]{Diagram of network used to include uncertainty estimates in the input vector. Rounded edged boxes represent inputs/outputs of network. Squared edges boxes represent one or more fully connected layers within the model where the internal model weights are optimised during training. The structure of these is described in Table \ref{net_config2}. The diamond box represents the concatenation of layers: this step involves no new weights and simply concatenates the outputs of the previous layers. The arrows represent the direction of flow of data through the network.}
	\label{fig:Uncer_MDN}
\end{figure}

Networks are trained for two different datasets: first a training set $T$ in which data were perturbed by 10\% Gaussian noise is used to train what we refer to herein as a \textit{Noisy-MDN}. This MDN does not include uncertainty in its input vector. A second training set $T_{uncer}$ includes a variable uncertainty vector as described in the previous section, which is used to train what we refer to as an \textit{Uncertainty-MDN}. To include both the dispersion curve and their uncertainties in the latter networks two inputs are included as shown in Figure \ref{fig:Uncer_MDN}. The dispersion curve is passed through 2 layers, whilst the uncertainty is input separately and passed through one layer. The outputs of these two layers is then concatenated before being passed through two more layers to output the parameter vector that defines the probability distribution of the shear wave velocity structure.

Separate MDNs are trained for each depth interval in the velocity structure defined in Figure \ref{fig:models_raw,models_sm}b. For each interval approximately 40 networks are trained from which we select for the ensemble the 10 networks with the lowest cost value across the validation set. The weights and biases are randomly initialized using the Glorot uniform initializer \citep{Glorot2010} for each training run, and we use different sizes of layers in the different networks to create diversity. The different layer sizes were determined using a form of Bayesian optimization using the Python library hyperopt \citep{Bergstra2015}. In Appendix A we describe the network configurations trained. The networks each use a Gaussian mixture with 15 kernels, so by using an ensemble of 10 networks a total of 150 kernels potentially contribute to each posterior distribution. However, we found that normally only 3 or 4 kernels with different means and standard deviations were assigned significantly non-zero amplitudes by each individual network.

\subsection{Network Evaluation}
\begin{figure}
	\begin{center}
	\includegraphics[width=\textwidth]{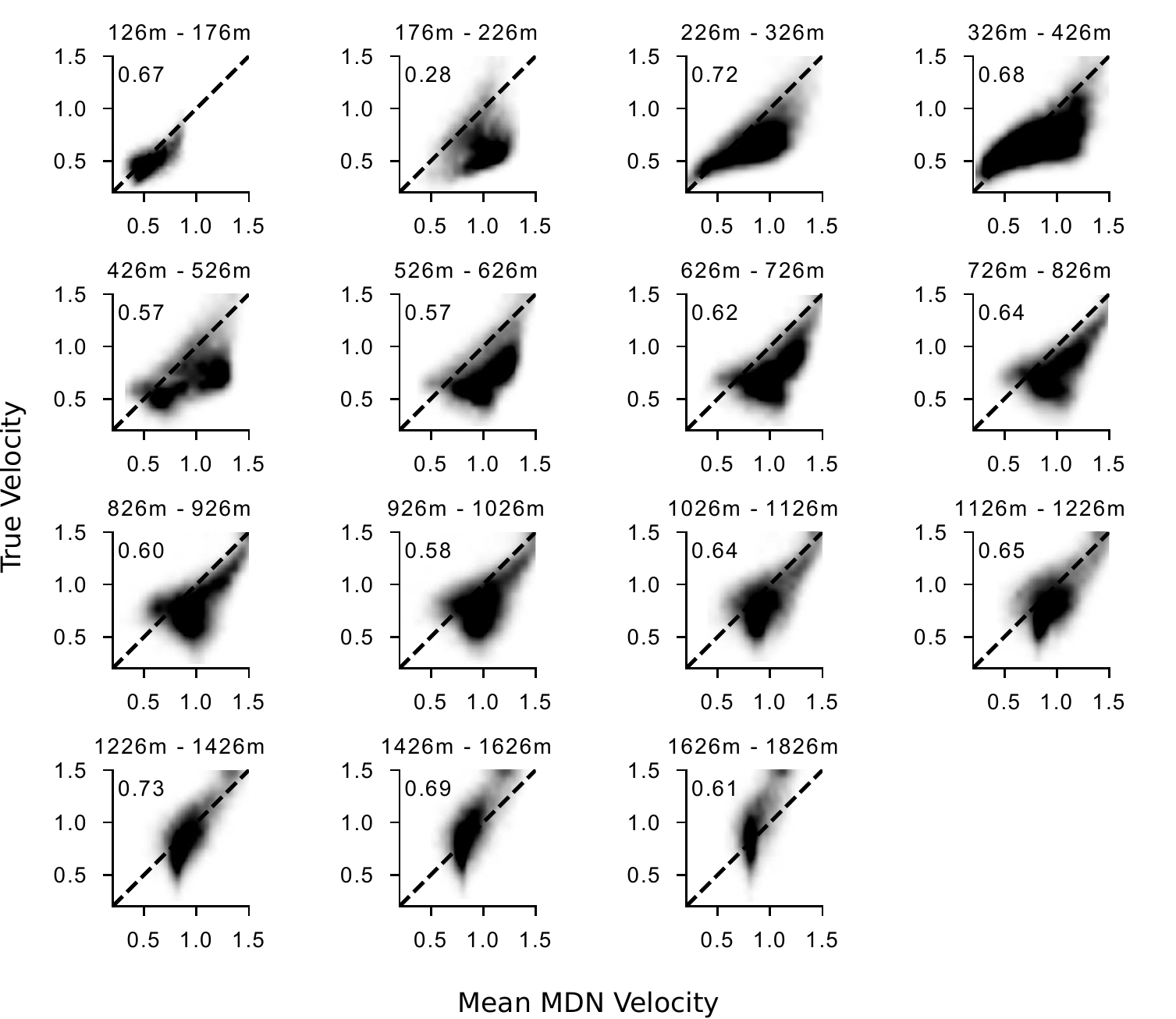}
	\end{center}
	\caption[Mean of the posterior marginal pdfs from Noisy-MDN inversions, versus the true value of velocity for each velocity structure in the set of smooth models]{Mean of the posterior marginal pdfs from Noisy-MDN inversions, versus the true value of velocity for each velocity structure in the set of smooth models. Each graph represents a different depth interval as indicated above the graph. The corresponding Pearson correlation coefficient R is given in the top left corner of each graph.}
	\label{fig:eval_fund_noise}
\end{figure}
\begin{figure}
	\begin{center}
	\includegraphics[width=\textwidth]{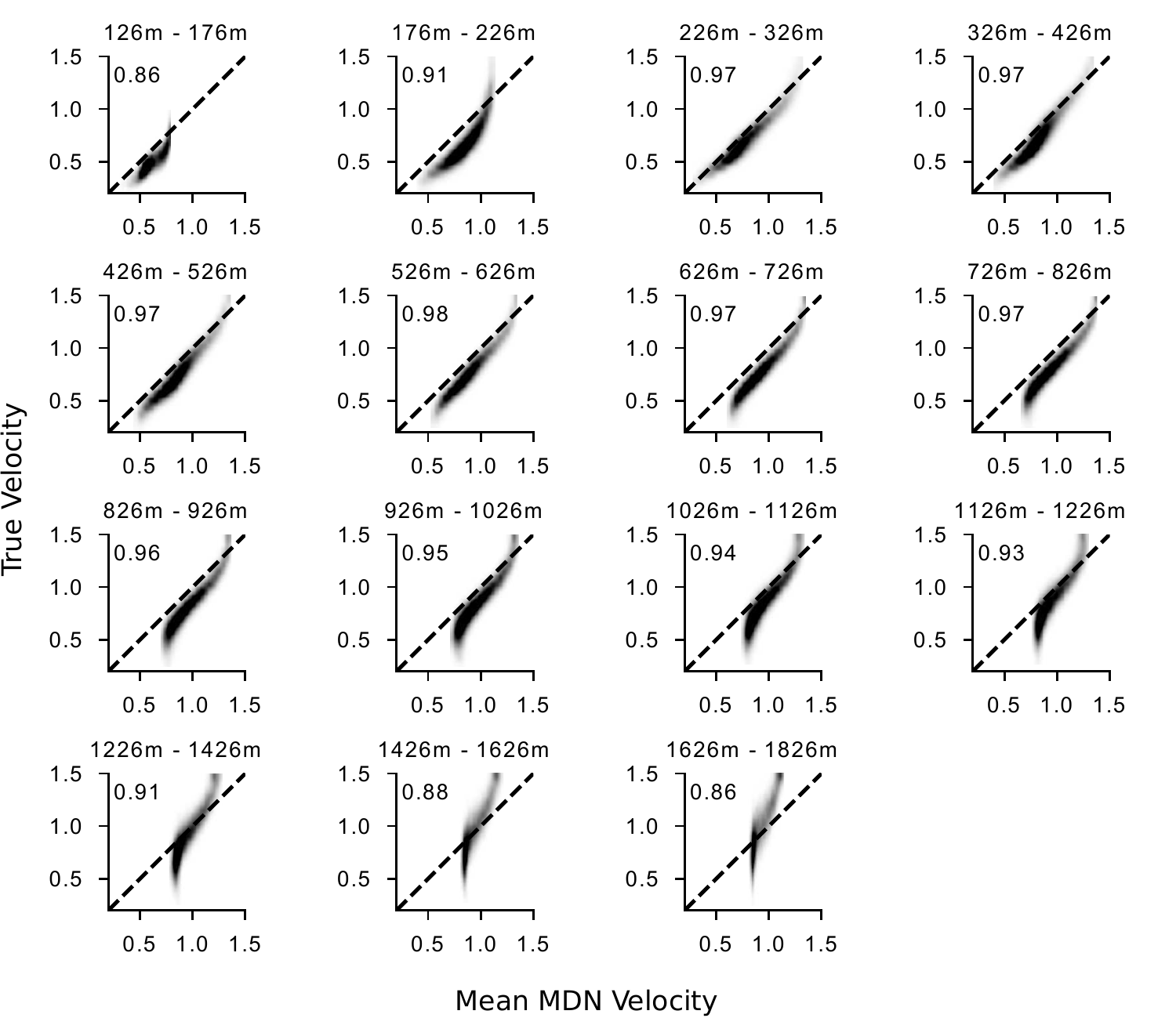}
	\end{center}
	\caption[Mean of the posterior marginal pdfs from Uncertainty-MDN inversions, versus the true value of velocity for each velocity structure in the set of smooth models]{Mean of the posterior marginal pdfs from Uncertainty-MDN inversions, versus the true value of velocity for each velocity structure in the set of smooth models. Each graph represents a different depth interval as indicated above the graph. The corresponding Pearson correlation coefficient R is given in the top left corner of each graph.}
	\label{fig:eval_uncer}
\end{figure}

A set of 100,000 synthetic velocity structures to which no network has previously been exposed were then created. These simulate relatively smooth velocity structures by not allowing the velocity to vary more than 400m/s between neighbouring depth intervals. Corresponding data are created using the DISPER80 forward modeller, to which 10\% Gaussian noise was added. For each depth interval in the velocity structure we apply the MDN ensemble to each of the 100,000 synthetic data and calculate the mean of each posterior marginal pdf $\bar{\mu}_{post}$
\begin{equation}\label{mean_post}
\bar{\mu}_{post} = \sum_{i=1}^cM\alpha_i\mu_i
\end{equation}

The correlation between the mean of the posterior and the true target value for each data vector can be used to evaluate the performance of the networks when presented with new data. This evaluation does not use all of the information contained in each posterior pdf, but does provide a practical way to begin to evaluate network performance. Figure \ref{fig:eval_fund_noise} shows the means of the posterior pdf of the fundamental mode Rayleigh wave Noisy-MDN inversions versus the true velocity values across all of the synthetic smooth velocity models, for each depth interval. The corresponding Pearson correlation coefficient, $R$, is shown in the top-left corner of the plot. The plots show a clear tendency for the mean of the network to over-estimate the true velocity value. When the same inversions are performed using the Uncertainty-MDNs (Figure \ref{fig:eval_uncer}) the correlation between the mean MDN velocities and the true velocities improves at every depth level. The additional information provided to the network that describes uncertainties in the data results in a significantly more accurate estimate of the velocity structure.

\begin{figure}
	\begin{center}
	\includegraphics[width=\textwidth]{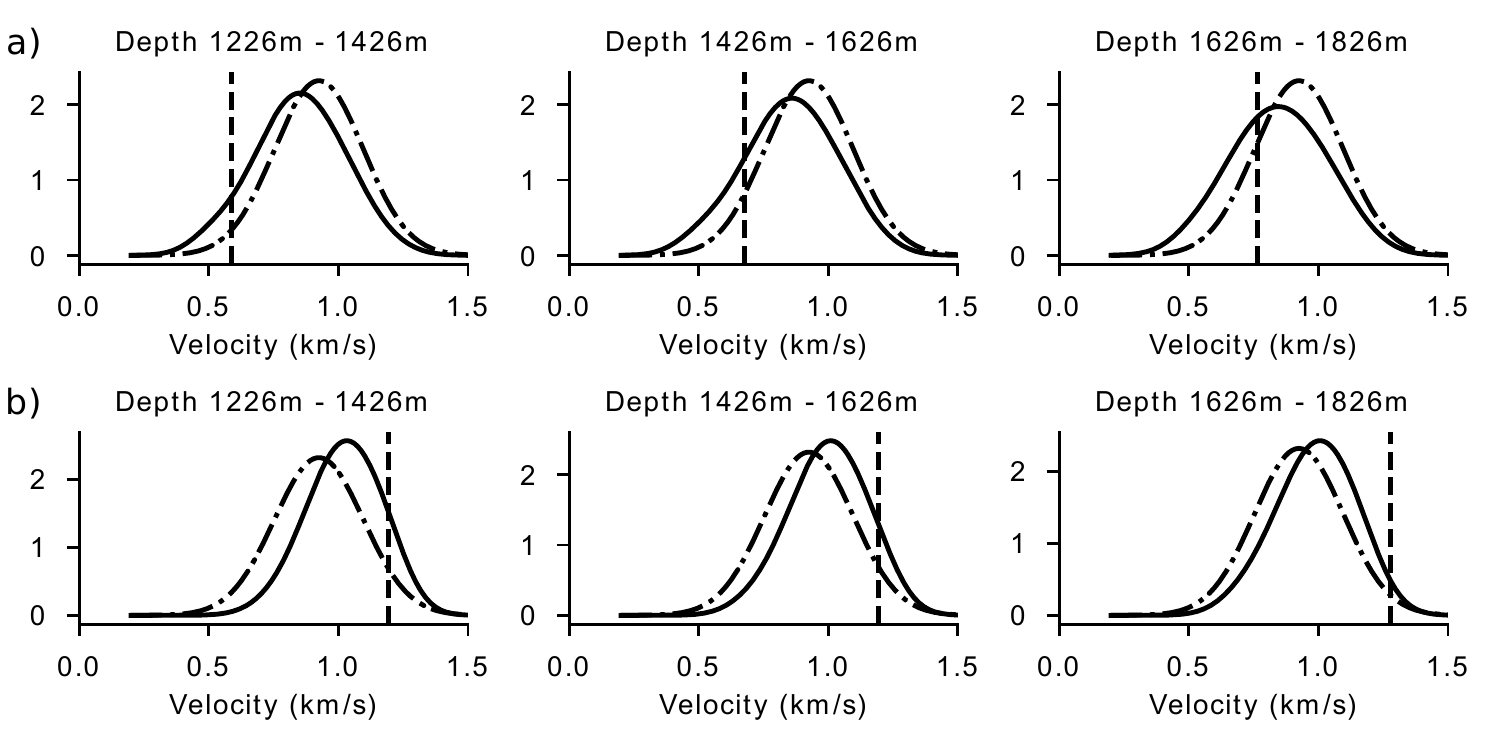}
	\end{center}
	\caption[Individual probability density functions for depths below 1226m for two synthetic velocity structures]{Individual probability density functions for depths below 1226m for two synthetic velocity structures in (a) and (b) respectively. The solid line is the marginal posterior probability density function from the MDN, the vertical dashed line is the true velocity value, and the dot-dash line is the prior probability density function.}
	\label{fig:prior_show}
\end{figure}

The plots in Figure \ref{fig:eval_uncer} allow us to compare how the networks perform at different depth levels. The performance of the networks decrease with depth, and at the deeper levels (1626-1826m) the mean of the Uncertainty-MDN tends towards the mean of the prior. Figure \ref{fig:prior_show} shows an example of the marginal posterior probability density function for two synthetic velocity structures at depths below 1226m. In both plots the true velocity structure is far away from the mean of the prior distribution yet the predicted marginal posterior distribution is very close to the prior: this shows that at these depths the networks are unable to add any information to the prior pdf given the data presented to the network. For this reason the following results are only shown down to a depth of 1226m.
\subsection{Synthetic  Results}
\begin{figure}
	\centering
		\includegraphics[width=\textwidth]{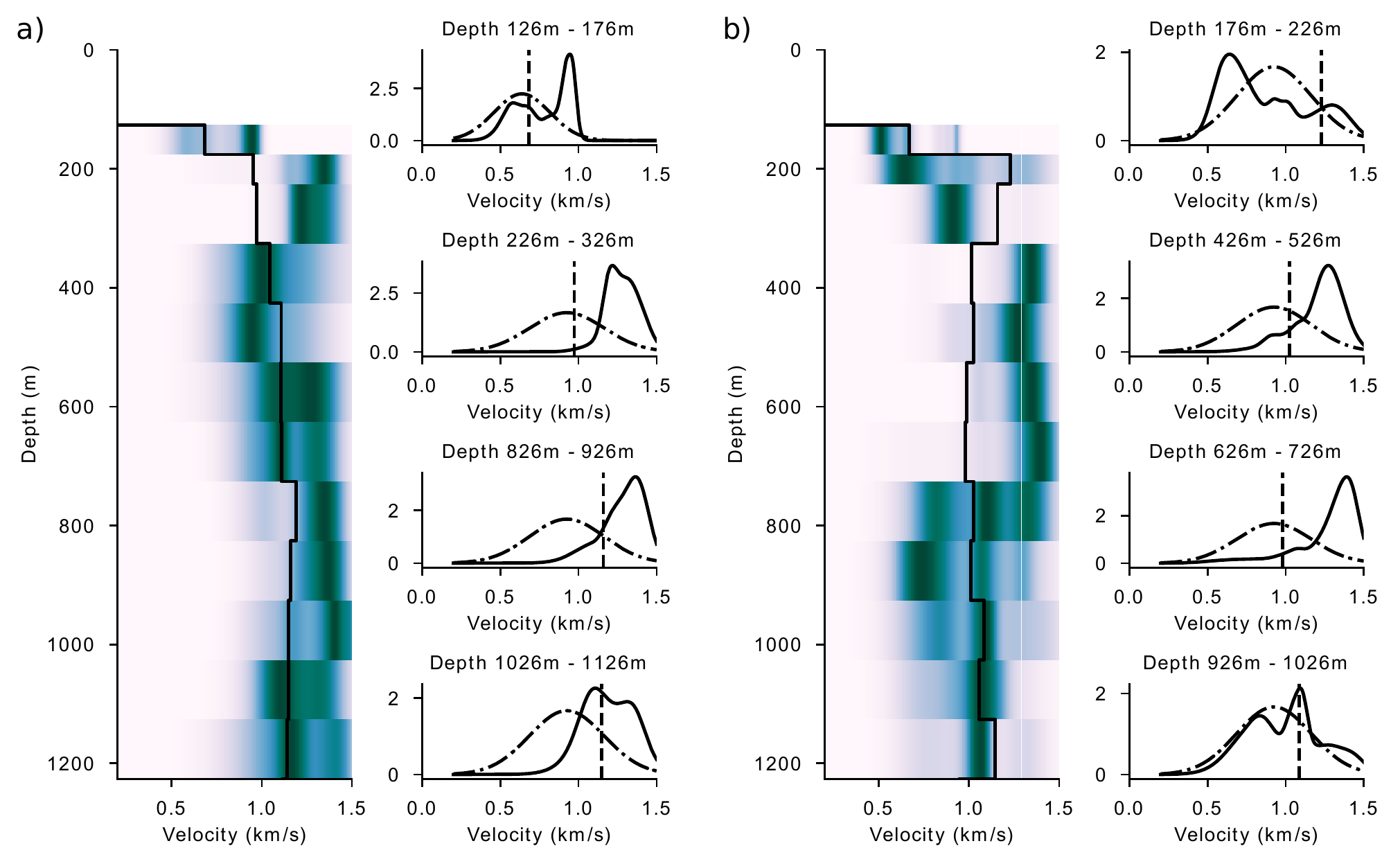}
	\caption[1D depth inversion result from Noisy-MDNs for two synthetic velocity structures with individual probability density functions shown for four depth levels]{1D depth inversion result from Noisy-MDNs for two synthetic velocity structures with individual probability density functions shown for four depth levels. In the depth inversions dark colours represent areas of higher probability, each row of the posterior integrates to unity, and the black solid line is the true synthetic velocity structure. In the individual probability density functions the solid line is the marginal posterior probability density function from the MDN, the vertical dashed line is the true velocity structure, and the dot-dash line is the prior probability density function.}
	\label{fig:synthetic_6657_noise,synthetic_37116_noise}
\end{figure}

\begin{figure}
	\centering
		\includegraphics[width=\textwidth]{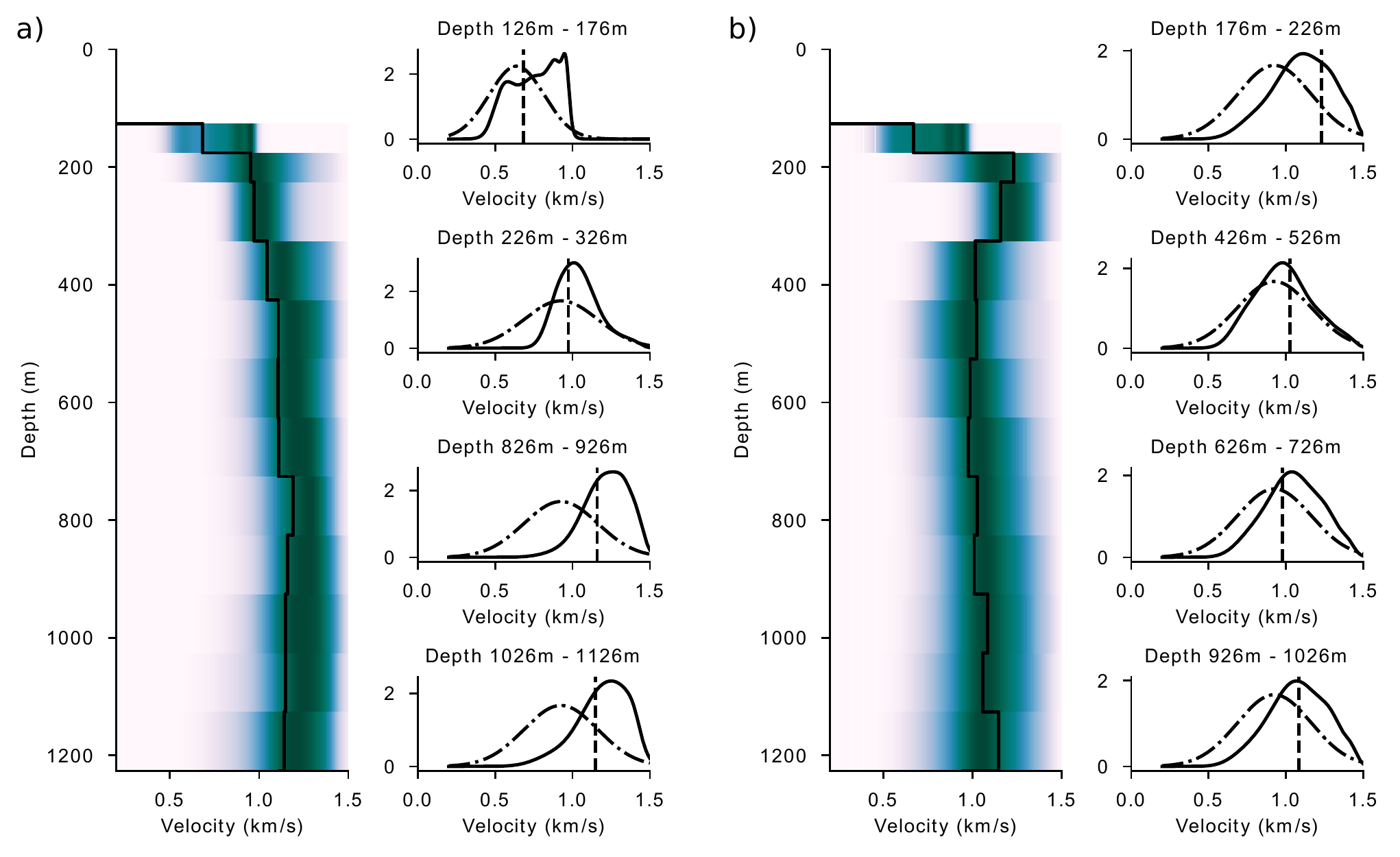}
	\caption[1D depth inversion result from Uncertainty-MDNs for two synthetic velocity structures with individual probability density functions shown for four depth levels]{1D depth inversion result from Uncertainty-MDNs for two synthetic velocity structures with individual probability density functions shown for four depth levels. In the depth inversions dark colours represent areas of higher probability, each row of the posterior integrates to unity, and the black solid line is the true synthetic velocity structure. In the individual probability density functions the solid line is the marginal posterior probability density function from the MDN, the vertical dashed line is the true velocity structure, and the dot-dash line is the prior probability density function.}
	\label{fig:synthetic_6657_uncer,synthetic_37116_uncer}
\end{figure}

Figure \ref{fig:synthetic_6657_noise,synthetic_37116_noise} shows the inversion of synthetic data from Noisy-MDN inversions: as seen in Figure \ref{fig:eval_fund_noise} the networks generally over-estimate the predicted velocity and often the uncertainty does not encompass the true solution. The addition of noise of fixed standard deviation to training data examples does not fully represent the true uncertainty of each individual data point, so the inversion of noisy data results in unreliable estimates of velocity. However, when adding the uncertainty estimates explicitly into the network, Uncertainty-MDN results produce more reliable estimates as shown by the 1D depth inversions in Figure \ref{fig:synthetic_6657_uncer,synthetic_37116_uncer}. The results from the networks are more representative of the true velocity structure, the maximum likelihood from the pdfs are much closer to the true shear wave velocity value, and the uncertainty ranges encompass the true velocity structure in all cases. This result is important for training and applying networks to field data. Noise, measurement error or assumptions about the Earth velocity structure can all affect the reliability of field data measurements; we never know whether our data is reliable, but we can often estimate the uncertainty in recorded data. Including the uncertainty information in networks allows them to be applied to field data with increased confidence that they will produce reliable estimates of posterior pdfs that encompass the true subsurface velocities.
\subsection{Field Data}
\begin{figure}
	\centering
		\includegraphics[width=0.7\textwidth]{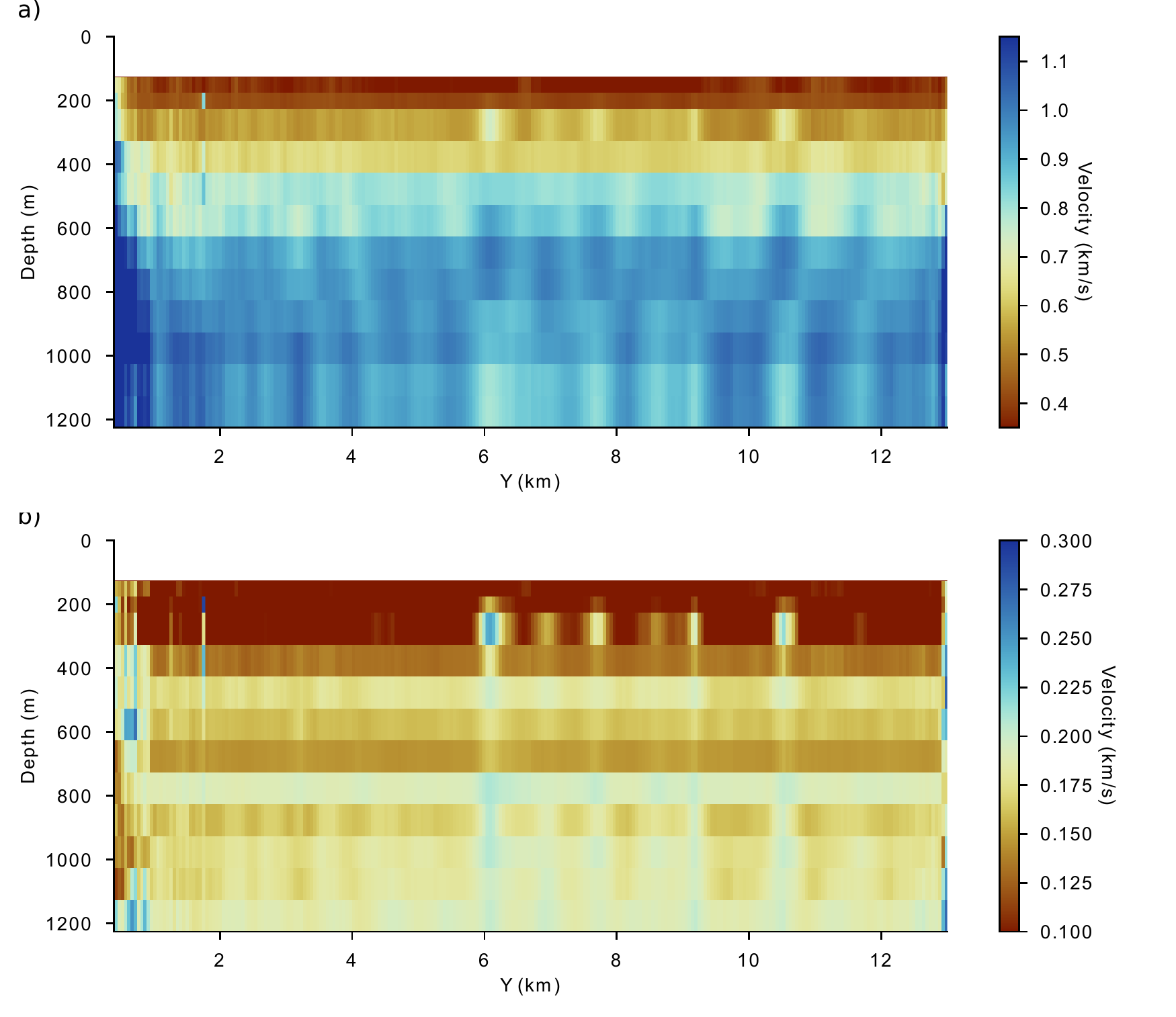}
	\caption[Mean shear velocity cross-section, and corresponding posterior standard deviation cross-sections from the Noisy-MDN inversion]{(a) Mean shear velocity cross-section, and (b) corresponding posterior standard deviation cross-sections from the Noisy-MDN inversion. The top white layer represents the water layer where shear velocity is zero.}
	\label{fig:Grane_mu_noise,Grane_std_noise}
\end{figure}
\begin{figure}
	\centering
		\includegraphics[width=0.7\textwidth]{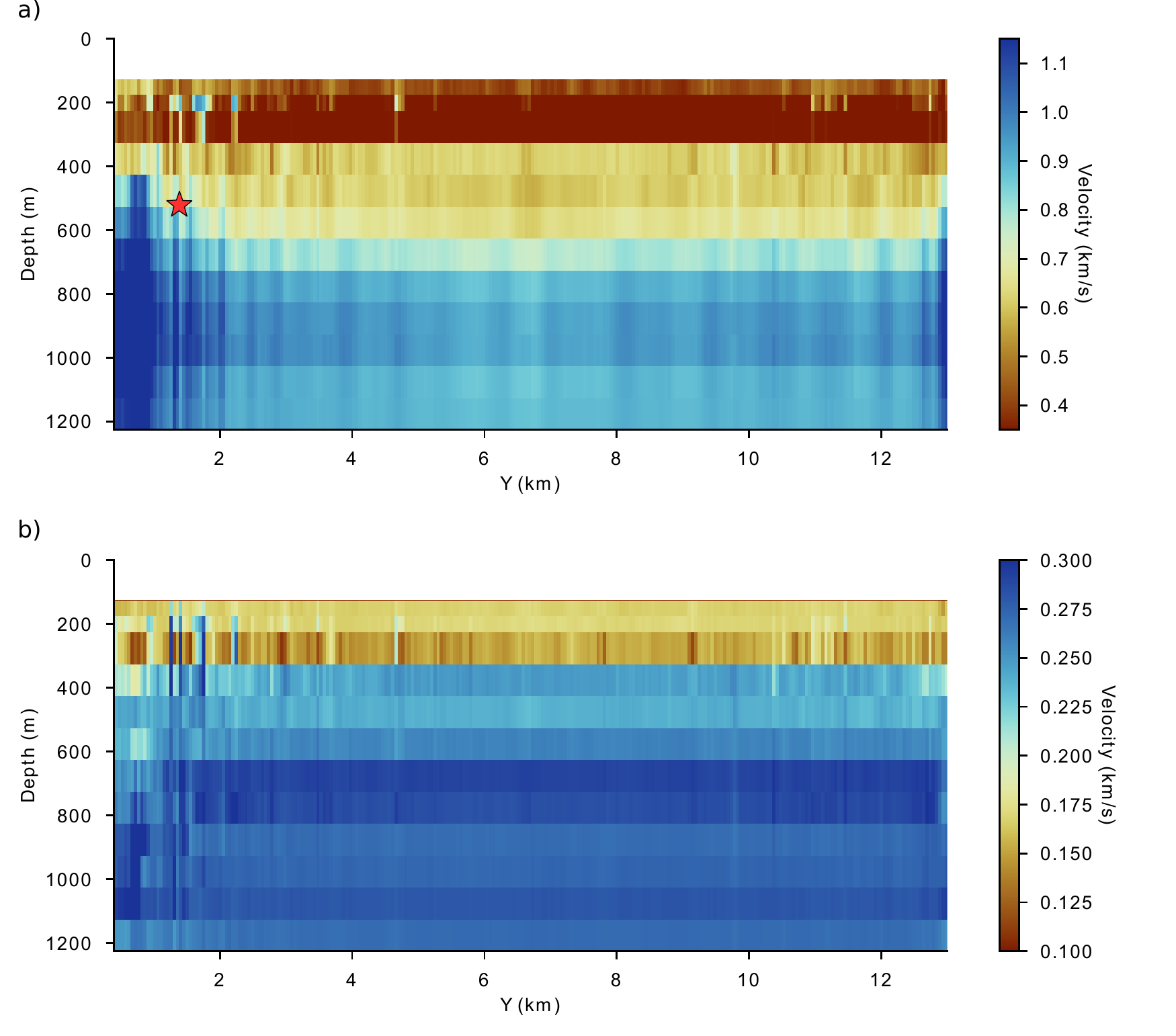}
	\caption[Mean shear velocity cross-section, and corresponding posterior standard deviation cross-sections from the Uncertainty-MDN inversion]{(a) Mean shear velocity cross-section, and (b) corresponding posterior standard deviation cross-sections from the Uncertainty-MDN inversion. The top white layer represents the water layer where shear velocity is zero. Red star shows location of the joint pdf shown in Figure \ref{fig:joint_pdfs}.}
	\label{fig:Grane_mu_uncer,Grane_std_uncer}
\end{figure}

The final trained MDNs are applied to invert Rayleigh wave phase velocities from the Grane field in the Norwegian North Sea. Dispersion curves were extracted at each grid point producing 26,772 dispersion curves to be inverted for 1D depth-velocity structures. The standard deviations shown in Figure \ref{fig:Phase_vel,Phase_uncer}b were extracted at each point and used as the uncertainty vector input to the Uncertainty MDNs (Figure \ref{fig:Uncer_MDN}). Figures \ref{fig:Grane_mu_noise,Grane_std_noise} and \ref{fig:Grane_mu_uncer,Grane_std_uncer} show the mean and associated standard deviations (representing uncertainty) of the posterior pdf estimated at the location of the black line in Figure \ref{fig:Phase_vel,Phase_uncer}a from Noisy-MDNs and Uncertainty-MDNs respectively. Both plots of the mean show a reasonably similar structure: a near-surface low velocity layer down to 300m, then an increased velocity down to 600m, with yet higher velocities below this. However, the layers are more distinct in the inversion using the Uncertainty-MDNs. Figure \ref{fig:Grane_mu_noise,Grane_std_noise}a from the Noisy-MDN shows a higher variability in the velocity below 600m than does the mean in Figure \ref{fig:Grane_mu_uncer,Grane_std_uncer}a, and the velocity highs in Figure \ref{fig:Grane_mu_noise,Grane_std_noise}a coincide with higher uncertainties in Figure \ref{fig:Grane_mu_noise,Grane_std_noise}b. When networks are trained including the full uncertainty information (Figure \ref{fig:Grane_mu_uncer,Grane_std_uncer}) these velocity highs disappear so that the mean velocity and uncertainties are laterally smoother across the section. We therefore now focus on the Uncertainty MDN results.

\begin{figure}
	\centering
		\includegraphics[width=\textwidth]{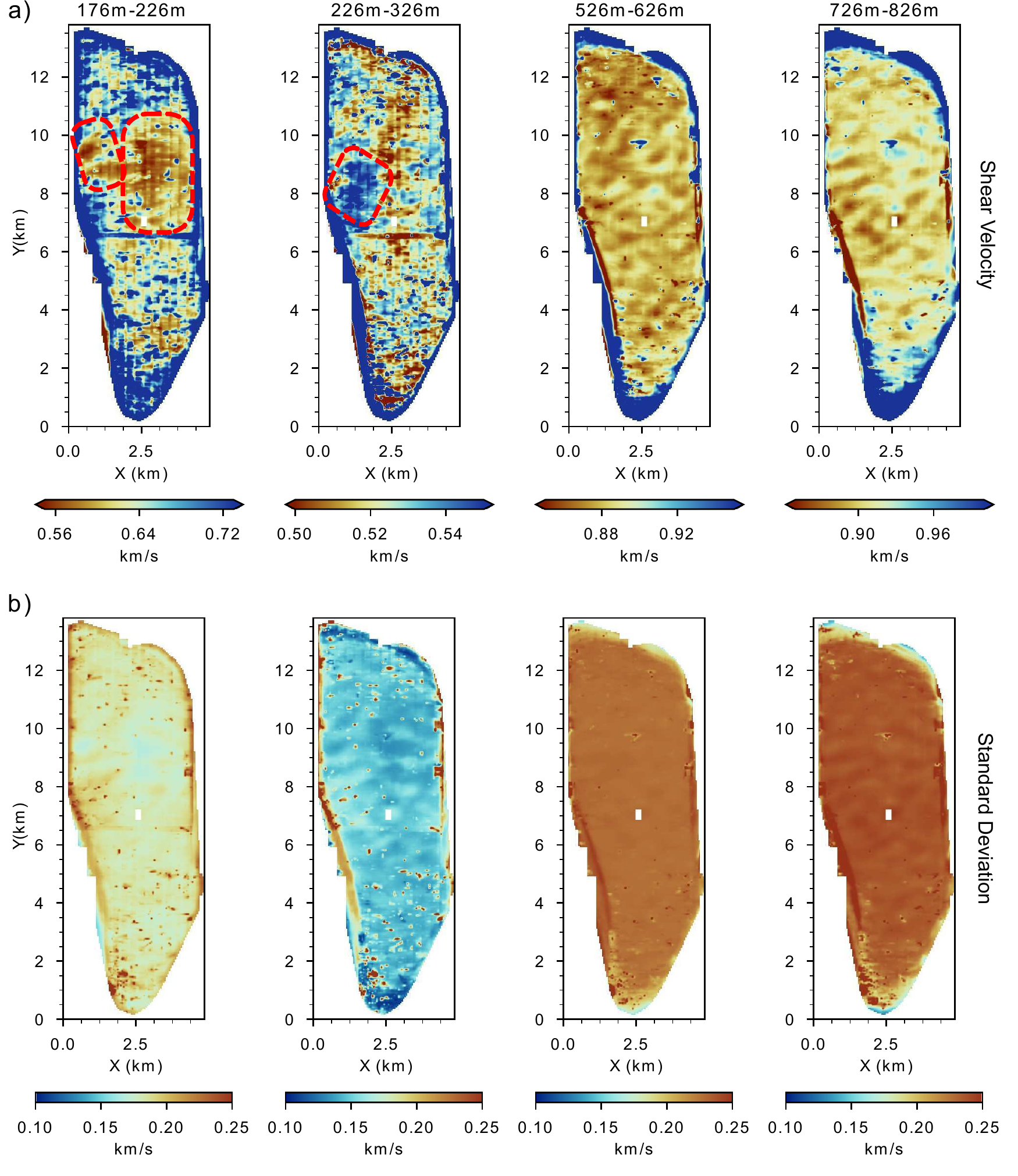}
	\caption[Fixed depth maps of the mean and the standard deviation of the shear velocity from Uncertainty-MDN inversion of fundamental mode Rayleigh dispersion]{Fixed depth maps of (a) the mean and (b) the standard deviation of the shear velocity from Uncertainty-MDN inversion of fundamental mode Rayleigh dispersion at depth slices 176m-226m, 226m-326m, 526m-626m, 726m-826m.}
	\label{fig:Grane_mean,Grane_std}
\end{figure}

Figure \ref{fig:Grane_mean,Grane_std} shows the mean and standard deviation horizontal depth slices from the Uncertainty-MDNs. In the near surface maps (126m - 326m) the results show similar structures to those in the phase velocity maps in Figure \ref{fig:Phase_vel,Phase_uncer}a at short periods, for examples within the dotted red boxes in Figure \ref{fig:Grane_mean,Grane_std}a. The deeper maps (536m - 826m) show structures similar to that of the longer period phase velocity maps, but also a higher standard deviation (Figure \ref{fig:Grane_mean,Grane_std}b) than shallower layers. As a result, the shear velocity variation in these deeper structures falls within their standard deviation, suggesting that they might not represent true structure. 

\begin{figure}
	\centering
		\includegraphics[width=0.7\textwidth]{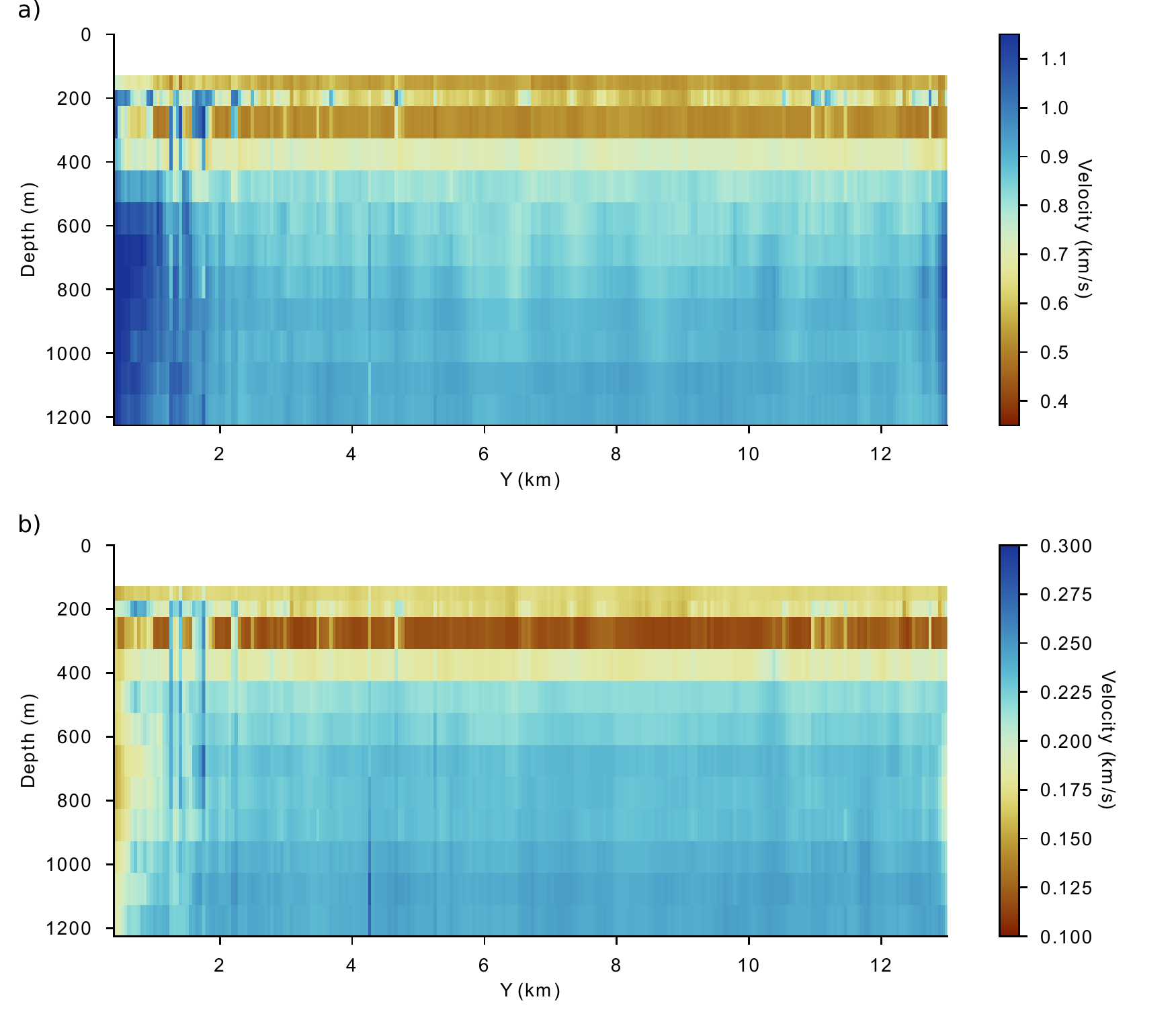}
	\caption[Mean shear velocity cross-section, and corresponding posterior standard deviation cross-sections from the Uncertainty-MDN inversion of fundamental and first higher mode Rayleigh dispersion]{(a) Mean shear velocity cross-section, and (b) corresponding posterior standard deviation cross-sections from the Uncertainty-MDN inversion of fundamental and first higher mode Rayleigh dispersion. The top white layer represents the water layer where shear velocity is zero.}
	\label{fig:Grane_mu_uncer_joint,Grane_std_uncer_joint}
\end{figure}
\begin{figure}
		\includegraphics[width=\textwidth]{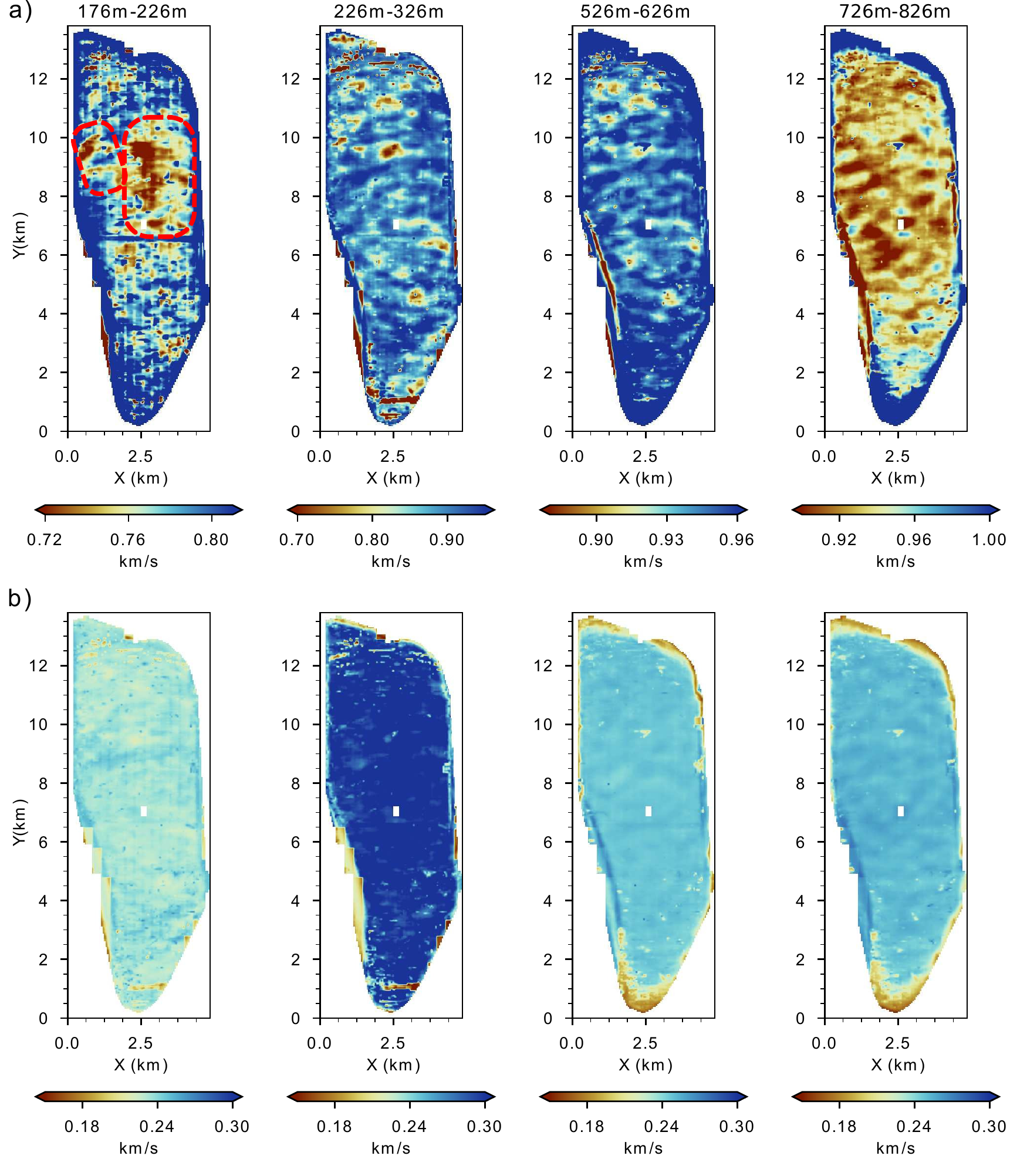}
	\caption[Fixed depth maps of the mean and the standard deviation of the shear velocity from Uncertainty-MDN inversion of fundamental and first higher mode Rayleigh dispersion]{Fixed depth maps of (a) the mean and (b) the standard deviation of the shear velocity from Uncertainty-MDN inversion of fundamental and first higher mode Rayleigh dispersion at depth slices 176m-226m, 226m-326m, 526m-626m, 726m-826m.}
	\label{fig:Graneh_mean,Graneh_std}
\end{figure}

The method outlined above can easily be extended to joint inversion of fundamental and first higher mode data by adding two further inputs: the vector of first higher mode phase velocity values generated from the velocity structures in the original training set and a vector of their associated uncertainties. 
Figure \ref{fig:Grane_mu_uncer_joint,Grane_std_uncer_joint} shows the cross-section results and Figure \ref{fig:Graneh_mean,Graneh_std} shows the results from 4 depths layers, 126m-176m, 226m-326m, 426m-526m, 626m-726m, from Uncertainty-MDN joint inversion. The same features seen in the shallow layer of Figure \ref{fig:Grane_mean,Grane_std}a are seen in the shallow layer of the joint inversion, highlighted by the red dashed boxes in Figure \ref{fig:Graneh_mean,Graneh_std}a. However, the velocities are on average higher than the fundamental mode-only MDN inversions and the standard deviations are larger. In addition, the depth slice at 226m-326m is entirely different to the corresponding slice in Figure \ref{fig:Grane_mean,Grane_std}a. Figure \ref{fig:Grane_mu_uncer_joint,Grane_std_uncer_joint}a shows that the low velocities observed in the top layers of the Uncertainity MDN fundamental model inversion (Figure \ref{fig:Grane_mu_uncer,Grane_std_uncer}a) no longer exist in the joint inversion with higher modes, showing the latter waves appear to have added additional information to the inversion. However, we are less confident about the quality of the higher mode dispersion measurements than those from the fundamental mode, so we include this result as a demonstration, but in the Discussion below we focus mainly on the fundamental mode results.
\section{DISCUSSION}\label{discussion}
We inverted Rayleigh wave phase dispersion curves for subsurface shear-wave velocities using MDN's trained with added Gaussian noise at a fixed standard deviation to simulate average data uncertainties, and a second type of MDN with the data uncertainty vector included as an additional input. We showed that to invert noisy data for reliable velocity structures the uncertainty estimates should be included in the network.

A constant number of fixed depth-velocity intervals were used in each MDN inversion, leading to inversions for effective medium (averaged) shear velocities for each fixed depth interval. A trans-dimensional network inversion would have had to include varying depths and number of layers which would significantly increase the dimensionality of the network inversion problem and require a much larger training set and more complex network structure. This in turn would increase training time and the memory needed for training, and would likely make the network outputs less stable and reliable since the posterior would effectively be emulating the inverse function in a higher dimensional space. For our intended application (to test our ability to rapidly monitor the overburden of a permanently instrumented field), the inversion for effective medium parameters over fixed depth intervals was sufficient. 
\subsection{Comparison with Monte Carlo Methods}
\begin{figure}
	\begin{center}
	\includegraphics[width=0.95\textwidth]{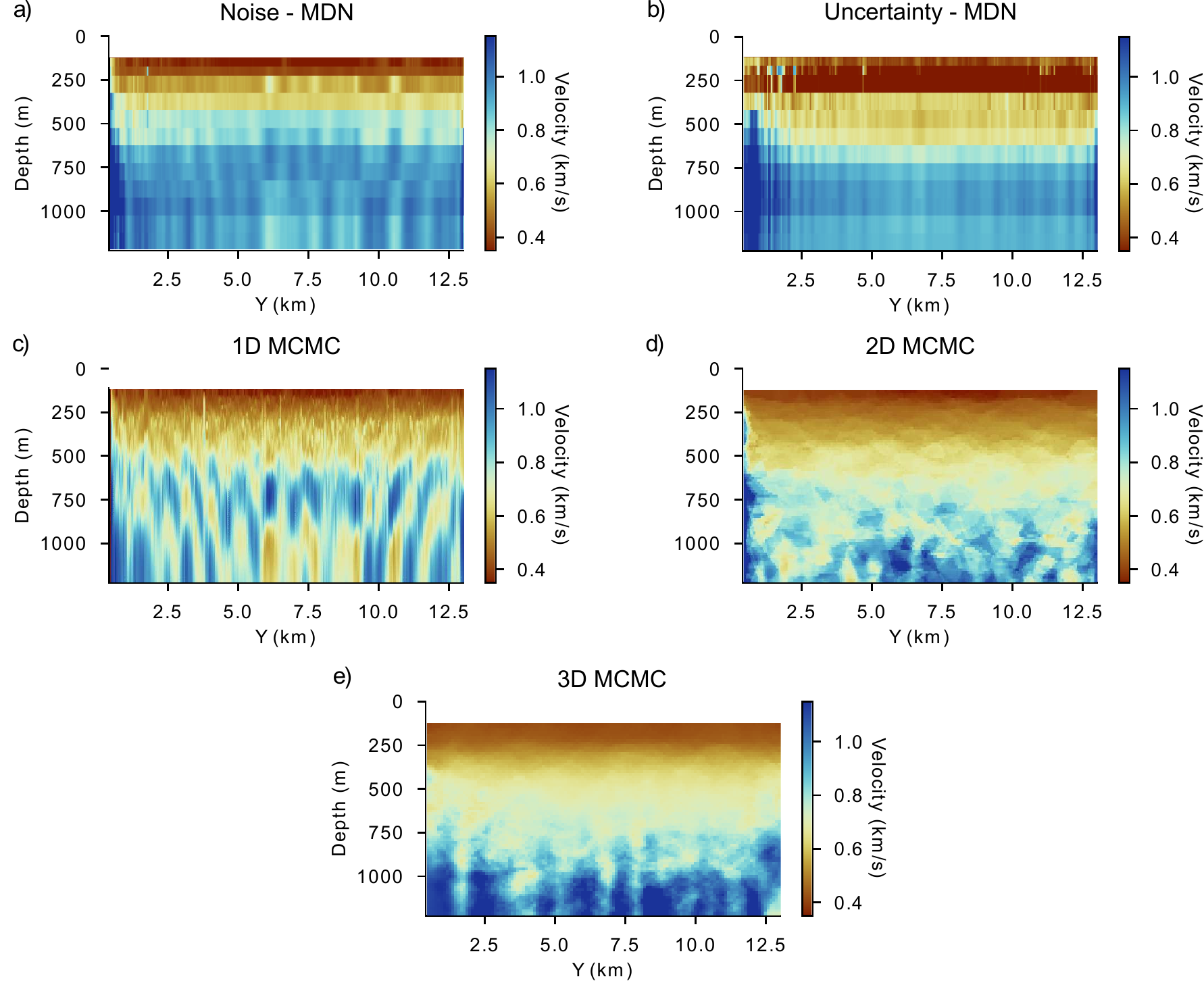}
	\end{center}
	\caption{Mean shear velocity along the cross-section in Figure \ref{fig:Phase_vel,Phase_uncer}a from (a) MDN inversions using a training set with added Gaussian noise of fixed standard deviation, (b) MDN inversions using estimated data uncertainties as added input data, (c) independent 1D Monte Carlo inversions, (d) a single 2D Monte Carlo inversion, and (e) a 3D Monte Carlo inversion, where results in (c), (d) and (e) are from \cite{Zhang2019}. The top white layer represents the water layer, where the shear velocity is zero.}
	\label{fig:MCMC_comp}
\end{figure}

\begin{table}
\caption{Table showing the mean-squared difference between the Noise- and Uncertainty-MDN inversion cross sections, and the Markov Chain Monte Carlo cross sections of \cite{Zhang2019}.}
\centering
\begin{tabular}{ccc} \toprule
 & Noisy-MDN & Uncertainty-MDN \\ \midrule
1D MCMC & 0.0018 & 0.0025 \\
2D MCMC & 0.0025 & 0.0026 \\
3D MCMC & 0.0021 & 0.0019 \\ \bottomrule
\end{tabular}
\label{tab:MSE_MCMC}
\end{table}
We compare the Noise- and Uncertainty-MDN inversion results to the Markov chain Monte Carlo results of \cite{Zhang2019}. Figure \ref{fig:MCMC_comp} shows the mean shear velocity cross sections of Figures \ref{fig:Grane_mu_noise,Grane_std_noise}a and Figures \ref{fig:Grane_mu_uncer,Grane_std_uncer}a along with the same cross sections from 1D, 2D and 3D trans-dimensional Markov chain Monte Carlo inversion (MCMC). Despite comparing a trans-dimensional result from Monte Carlo methods with fixed-depth layer results from MDNs, all cross sections show a similar, approximately 3-layered structure. The 1D MCMC (Figure \ref{fig:MCMC_comp}c) most represents the networks trained using the Noisy-MDNs (Figure \ref{fig:MCMC_comp}a) as both contain vertical velocity anomalies in the deeper part of the section. The Uncertainty-MDN has smoother variations laterally but also has a thicker near surface velocity layer and the second layer extends deeper into the section (to $\sim$700m); this is more representative of the 2D and 3D MCMC results (Figure \ref{fig:MCMC_comp}d and \ref{fig:MCMC_comp}e). This is confirmed by examining the mean-squared difference (MSD) between the mean of each MDN inversion and the Monte Carlo inversions in Table \ref{tab:MSE_MCMC}: the Noisy-MDN has a lower MSD compared to the 1D MCMC inversion and the uncertainty MDN has a lower MSD compared to the 3D MCMC inversion. This implies that by adding uncertainties to the MDN training we allow smoothness in the mean estimates which the 3D MCMC results suggest is reasonable across Grane.

\subsection{Joint Posterior Probability Density Functions}
The results in the previous section are created from the 1D marginal posterior pdf $p( m^i\mid \textbf{d})$ of the shear velocity in each layer independent of other velocities in each 1D profile. The correlations between velocities at different depths cannot be derived from such results. To estimate correlations it is necessary to analyse the joint posterior pdf $p( m^i, m^{i+1}\mid\textbf{d})$, which can be constructed from the product of the conditional and marginal pdfs.
\begin{equation}\label{joint_cond}
p( m^i, m^{i+1}\mid\textbf{d})=p( m^{i}\mid \textbf{d}) \times p(m^{i+1}\mid m^{i},\textbf{d})
\end{equation}
The marginal pdfs $p( m^{i}\mid \textbf{d})$ are given by the results shown in the previous sections. New networks are trained to estimate the conditional pdfs $p( m^{i+1}\mid m^{i},\textbf{d})$ by extending the input vector of the data with the velocity to which we want to condition our data: in this example this is the velocity of the layer above the one being estimated.
\begin{figure}
	\begin{center}
	\includegraphics[width=0.4\textwidth]{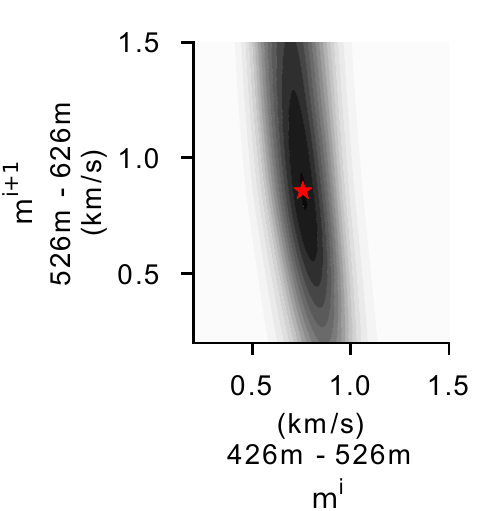}
	\end{center}
	\caption[Joint pdf comparing the velocity trade-off between two adjacent layers $m^i$ and $m^{i+1}$]{Joint pdf comparing the velocity trade-off between two adjacent layers $m^i$ and $m^{i+1}$ at depths given in the axis labels. The red star represents the mean velocity shown in Figure \ref{fig:Grane_mu_uncer,Grane_std_uncer}a.}
	\vspace{15pt}
	\label{fig:joint_pdfs}
\end{figure}

Figure \ref{fig:joint_pdfs} shows the results from the location shown by the red star in the Grane cross-section from Figure \ref{fig:Grane_mu_uncer,Grane_std_uncer}a. The plot shows a weak negative correlation, representing the weak trade-off between velocities in subsequent layers. This is likely to be because a relatively coarse parametrisation (compare that in Figure \ref{fig:models_raw,models_sm}a and \ref{fig:models_raw,models_sm}b) was used over depth for the inverse problem. If a finer parametrisation was used, such trade-offs would emerge more strongly as demonstrated by \cite{Meier2007}.
\subsection{Inversion Speed}
Post-training, neural networks invert new data extremely rapidly: in this study it took approximately 21 CPU seconds to invert all 26,772 locations. The results are compared to Monte Carlo methods which are known to be computationally expensive \citep{Bodin2009}: the MCMC methods used to create the crosslines shown in Figure \ref{fig:MCMC_comp} took approximately 186 CPU hours for 1D, 206 CPU hours for 2D, and 4824 CPU hours for 3D inversions. Despite the higher vertical resolution of results from MCMC methods (since the parametrisation over depth varies in those inversions), the compute-time for inversions is between 4 and 6 orders of magnitude larger than for trained MDNs.
\begin{figure}
	\begin{center}
	\includegraphics[width=0.95\textwidth]{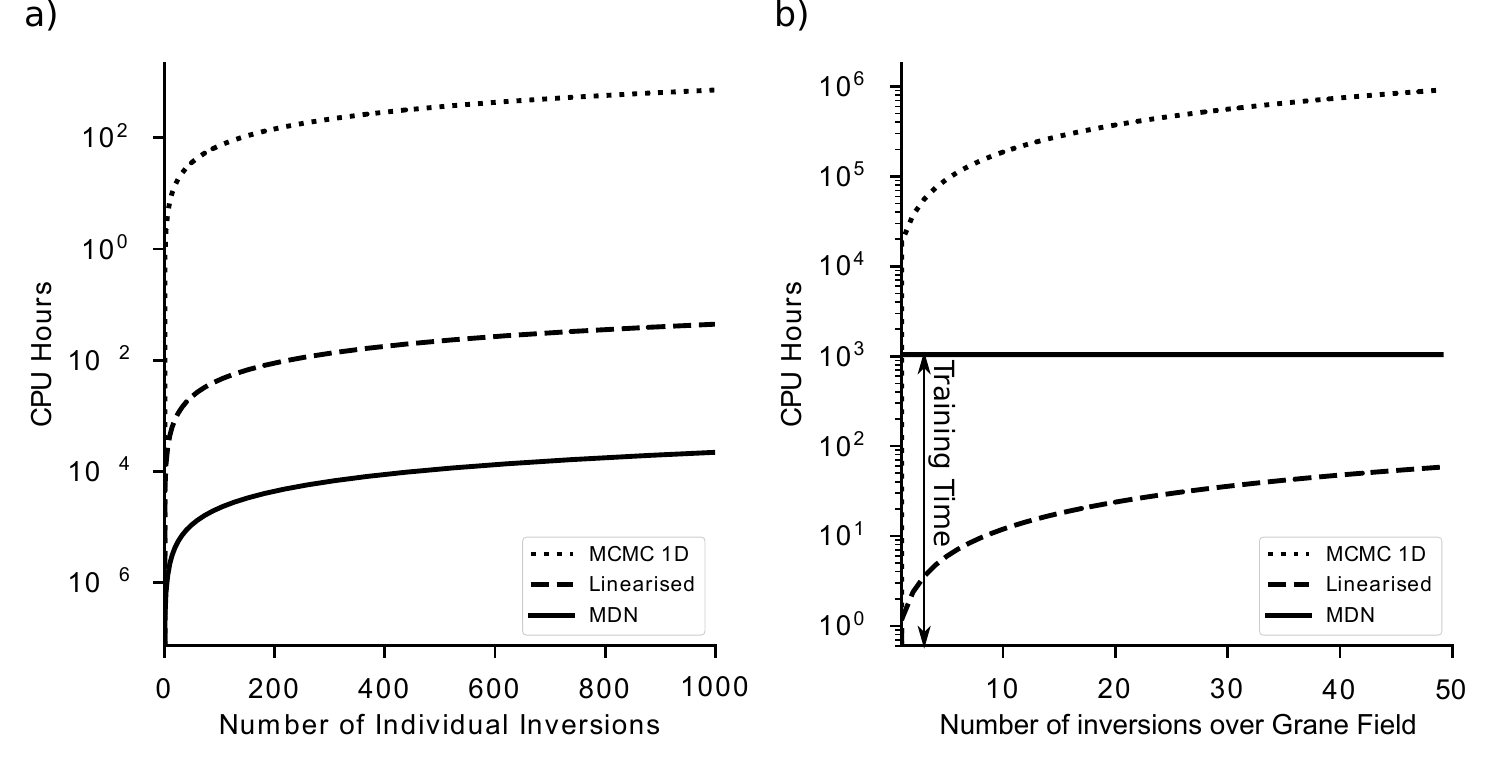}
	\end{center}
	\caption{Plots showing the CPU hour time for (a) one inversion per location and (b) the inversion over the entire Grane field using MDNs, linearized inversion and Monte Carlo 1D (MCMC 1D)  methods. The typical time taken to train the MDN (including the forward modelling runs) is included in (b) but not in (a).}
	\label{fig:timings}
\end{figure}

A comparison of time per inversion of an individual location for 1D MDN, 1D Monte Carlo and a 1D linearized inversion is shown in Figure \ref{fig:timings}a. Monte Carlo inversion is computationally the most expensive, and MDN inversion is two orders of magnitude faster than even linearized inversion. However, in this comparison we only accounted for the speed of the inversion which is not the full computational expense involved in using neural networks. Training a network before the inversion takes significant computational time: in this study we took 1280 CPU hours to train all of the networks used. It should be noted that training the network needs only to be done once, and hence is independent of number of locations to be inverted; therefore inverting more locations renders the MDN inversion method more computationally efficient. Figure \ref{fig:timings}b compares the CPU hours needed for monitoring-style repeated inversions across the full Grane field as performed in this study, including the time required for training MDNs. The initial cost of training a network before the first inversion is high, but thereafter repeated inversion of new data sets is nearly free. In comparison to 1D MCMC methods, even accounting for the initial training period, neural network methods are more efficient. Linear inversion methods are computationally cheaper that MDN methods: in this case approximately 1000 inversions of the same field would be needed for the neural network method to become cheaper. Surface wave tomography is a non-linear inversion problem and despite the linearized inversions involving fewer CPU hours they can only give approximate solutions, in particular for uncertainty estimates, due to their implicit assumption of incorrect (linear) physics. The neural network method provides a fully probabilistic, fully non-linear solution that, once a network is trained, can be used to obtain rapid, repeated inversions.
\section{CONCLUSION}\label{conclusion}
We trained mixture density networks (MDN's) to invert fundamental mode Rayleigh wave dispersion curves for subsurface shear-wave velocity using two different methods to represent data uncertainties. The MDNs give a fully probabilistic solution to this non-linear inversion problem giving comparable results to Monte Carlo solutions. We show that inputting data uncertainties explicitly to the network provides a more reliable solution estimate on noisy synthetic data, and a smoother result that is more similar to 3D Monte Carlo inversion results on field data. The same method is used for joint inversion of fundamental and first higher mode data. Once trained, the neural network approach gives rapid results that can be repeatedly applied to similar types of data in monitoring scenarios.
\section*{Acknowledgements}
The authors would like to thank the Grane license partners Equinor ASA, Petoro AS, ExxonMobil E\&P Norway AS and ConocoPhillips Skandinavia AS for allowing us to publish this work. The views expressed in this paper are the views of the authors and do not necessarily reflect the views of the license partners. The authors thank the Edinburgh Interferometry Project sponsors (Equinor, Schlumberger Cambridge Research and Total) for supporting this research.
\section*{Appendix 1: Network configurations}\label{Append1}
The terminology used here is standard for neural networks and is defined succiently in \cite{Bishop1995}. The networks using Gaussian noise to simulate uncertainty in the data were trained using 3 fully connected layers (FC), where each node receives an input from every node in the previous layer. Between each node of the FC layers a rectified linear unit (ReLU) is used. The individual layer sizes and the total number of parameters to be trained in each network is outlined in Table \ref{net_config}.
\begin{table}
\caption{Network configurations of the networks for which Gaussian noise of fixed standard deviation was added to the training set. Each network structure is trained 5 times with different random initialisations of starting parameter values.}
\centering
\begin{tabular}{cccc} \toprule
FC 1 & FC 2 & FC 3 & Total parameters \\ \midrule
 200            & 300           & 200        &  133,145\\
400            & 200            & 350        &   173,545 \\
400           & 1000           & 150        &   565,145\\
200            & 1000          &350         &   570,745\\
400            & 500            & 350        &    398,845\\
400           & 1000            & 200       &   617,445\\
300            & 300            & 150        &  147,645\\
400            & 1000            & 350      &   774,345\\ \bottomrule
\end{tabular}
\label{net_config}
\end{table}

The networks that included uncertainties as inputs were trained using 2 fully connected layers connected to the dispersion curve data and one fully connected layer connected to the uncertainty data, before concatenating the layers together and applying a further two hidden layers of size 250 and 150 respectively (Figure \ref{fig:Uncer_MDN}). In between each node of the fully connected layers a rectified linear unit (ReLU) is used. The individual layer sizes and the total number of parameters to be trained in each network is outlined in Table \ref{net_config2}.
\begin{table}
\caption{Network configurations of the networks that included uncertainty vectors in the training set. Each network structure is trained 5 times with different random initialisations of starting parameter values.}
\centering
\begin{tabular}{cccc} \toprule
\multicolumn{2}{c}{Dispersion} & Uncertainty & Total \\
FC 1 & FC 2 & FC 3 & parameters \\ \midrule
1295            & 240           & 500        & 573,045 \\
1100            & 900            & 550        &    1,427,795\\
960           & 860           & 400       &   1,210,635\\
1000            & 220          &1000         &   605,915\\
950           & 1000            & 140        &    1,300,315\\
1100           & 800          & 450       &   1,265,895\\
930            & 960            & 100        &  1,221,995\\
1200            & 200            & 600      &   517,295\\ \bottomrule
\end{tabular}
\label{net_config2}
\end{table}

\bibliographystyle{natbib}  
\bibliography{EIP_Report_NN} 

\end{document}